\documentclass[journal=jcpl,manuscript=letter]{achemso}

\usepackage{chemformula} 
\usepackage[T1]{fontenc} 
\usepackage[textsize=tiny]{todonotes}

\author{Inon Sharony}
\affiliation{School of Chemistry, Tel Aviv University, Tel Aviv 69978, Israel}
\author{Renai Chen}%
\affiliation{Department of Chemistry, University of Pennsylvania, Philadelphia, PA  19104, USA}
\author{Abraham Nitzan}
\affiliation{School of Chemistry, Tel Aviv University, Tel Aviv 69978, Israel}
\alsoaffiliation{Department of Chemistry, University of Pennsylvania, Philadelphia, PA  19104, USA}
\email{anitzan@sas.upenn.edu}

\title[]
  {Stochastic Simulation of Nonequilibrium Heat Conduction in Extended Molecule Junctions}

\abbreviations{SMJ,MM,MD,NEMD,NESS}

\begin{document}

\begin{tocentry}





\begin{figure}[H]
    \centering
    \includegraphics[width=1\textwidth]{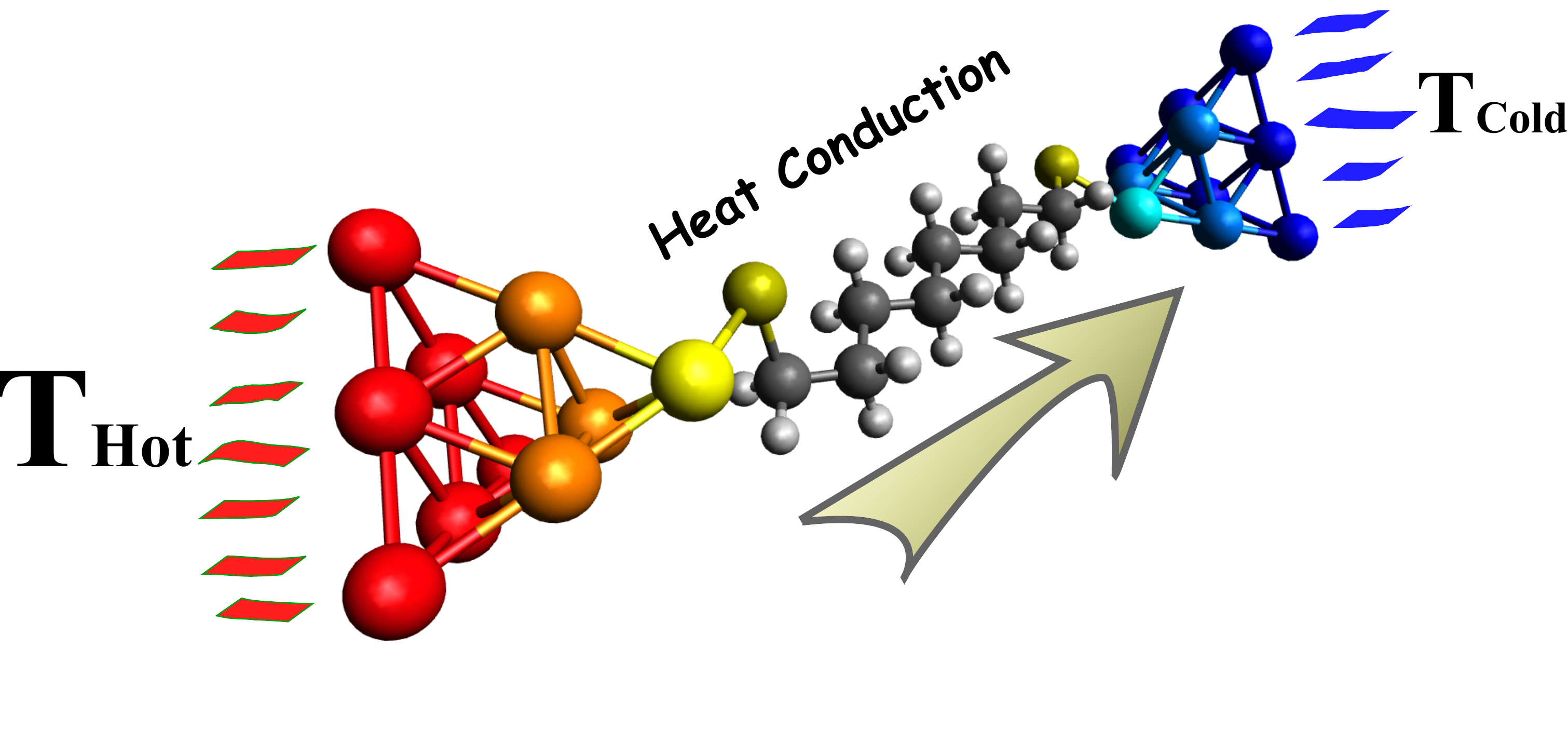}
    \caption{}
\end{figure}

\end{tocentry}

\begin{abstract}
Understanding phononic heat transport processes in molecular junctions is a central issue 
in the developing field of nanoscale heat conduction and manipulation. 
Here we present a Stochastic Nonequlibrium Molecular Dynamics simulation framework 
to investigate heat transport processes in molecular junctions in and beyond the linear response regime. 
We use extended molecular models which filter Markovian heat reservoirs through an intermediate substrate region, 
to provide a realistic and controllable effective bath spectral density. 
The results obtained for alkanedithol molecules connecting gold substrates 
agree with previous nonequilibrium Green's function calculations in frequency domain, 
and match recent experimental measurements 
(e.g. thermal conductance around 20 pW/K for alkanedithiols in single molecular junctions) 
Classical MD simulations using the full molecular forcefield 
and quantum Landauer-type calculations based on the harmonic part of the same forcefield are compared, 
and the similarity of the results indicate that heat transport is dominated by modes in the lower frequency range. 
Heat conductance simulations on polyynes of different lengths illuminates the effects of molecular conjugation on thermal transport. 
\end{abstract}

\section{Introduction}

Heat conduction in molecular junctions\cite{Segal2016arpc} 
has become a subject of increasing interests as a sub-field 
of nanoscale energy dissipation and transport over the last decade\cite{Pop2010}, 
driven by technological considerations of both stability and functionality 
of envisioned molecular electronic devices 
as well as the need to understand the fundamentals of heat transport in nanosize systems\cite{Li2012rmp,Dubi2011}. 
The most common approach to such calculations is based on classical MD simulations 
with substrate temperature controlled by generalized Langevin baths, 
with the obvious deficiency of misrepresenting the dynamics of high frequency modes, 
relying on the assumption that molecular heat conduction is dominated by modes in the lower frequency regime. 

Alternatively, quantum calculations were done, 
mostly based on the non-equilibrium Green's Function (NEGF) \cite{Yamamoto2006,Wang2006prb} methodology 
usually using the harmonic part of the molecular force field, 
leading to Landauer-type expressions\cite{landauer1957} for the molecular heat conduction 
in the harmonic approximation analogous to its counterpart 
in the problem of molecular electronic transport using free electron models. 
Given the different ranges of applicability of classical dynamics on the one hand 
and harmonic quantum dynamics on the other, 
comparing their performance in evaluating and predicting heat molecular conduction is obviously of interest.

Another powerful tool to investigate many-body interactions, without necessitating the harmonic approximation, is using atomistic Molecular Dynamics (MD) simulations.
MD simulation of an ergodic system allows calculation of Statistical Mechanical properties of a system (e.g. thermal conductivity) by analysis of the atomic trajectories.
MD simulations from previous reports, however, are mostly focused on specific systems such as liquids\cite{Zhang2005jpcb}, thin-films\cite{Lukes2000jht}, Graphene\cite{Berber2000prl}, or one-dimensional metal/semi-metal chains or wires\cite{Ness2017jcp}.
It is unclear how applicable these system-tailored simulations are to the thermal conduction in Single Molecule Junctions (SMJ).
A fully-functional MD simulation tool to study the structural dependence of molecular heat conduction, in which a full force-field is applied without particular system restrictions, is still lacking.

Focusing on classical simulations, 
equilibrium MD (EMD) is one of the easiest approaches to implement. 
One essentially applies the Green-Kubo formula 
to the time-autocorrelation of the current 
to get the thermal conductivity in the linear response regime
\cite{Ladd1986prb,Volz2000prb,Che2000jcp,Schelling2002prb,
McGaughey2004ijhmt1,McGaughey2004ijhmt2,Chen2012jap,Sellan2010prb}. 
Besides its limitation to linear response approximation, 
the method also suffers from slow convergence and limited applicability to heterogeneous systems\cite{Schelling2002prb}. 
Alternatively, under the nonequilibrium MD (NEMD) approach
\cite{Baranyai1996, Poetzsch1994, Oligschleger1999,Berber2000prl,Schelling2002prb, Jiang2010jap} 
one creates a temperature gradient by separating the simulated system into "slabs", 
and rescaling the atomic velocities at the "heat source" and “sink” slabs 
to set the temperature boundary conditions. 
In implementing this methodology care has to be taken 
for the finite-size effects associated with the so-imposed boundary conditions\cite{Schelling2002prb,Sellan2010prb}. 
Plus, the thermal bath effects are relatively obscure for this method. 
Another popular tool is the so-called Reversed Nonequilibrium MD (RNEMD)
\cite{Jund1999,Muller-Plathe1997,Muller-Plathe1999,Zhang2005jpcb,Bagri2011nl,Dong2014scirep,Tang2013apl,Si2017ijhmt}
in which the effect (fluxes) and the cause (temperatures) are reversed: 
one creates temperature differences by separating the simulated system into "slabs", 
and enforces a given heat flux on the system 
by taking a certain amount of kinetic energy from the "heat source" slab 
and put it into the "heat sink" slab, 
until the system reaches steady state at which the temperature at the source 
and sink sides is determined. 
Since in this approach non-equilibrium is imposed by a constant heat flux, 
it is limited to steady-state calculations. 

The stochastic nonequilibrium MD (SNEMD) methodology 
used in the present work is a variant of the NEMD outlined above, 
in which velocity rescaling reflects the interaction with a generic (white) thermal bath 
in a way consistent with the fluctuation-dissipation theorem. 
To account substrate actual spectral properties a section adjacent to the molecule is modeled explicitly 
and filters the effects of the generic stochastic dynamics\cite{Goga2012jctc} 
applied to bulk layers further from the molecular bridge. 
We show the stability and applicability of our method 
in calculating the temperature distribution, heat current, and thermal conductance 
in various molecular junction settings. Under our approach, 
the concept of temperature and build-up of thermal bias come in naturally, 
without manually perturbing the system at each simulated time-step or reversing causality.

This paper is the first in a series in which 
we plan to study the interplay between molecular composition and structure 
and its heat transport properties. 
For this purpose we have developed a numerical tool (described below) 
that can be readily adapted to different molecules and structures. 
Here we apply our tool to the study of heat transport in single alkane chains, 
a system that has been studied numerically\cite{Segal2003,Kloeckner2016} 
and experimentally (mostly for alkane layers\cite{Wang2006apl,Meier2014prl,Majumdar2015nl}, 
but very recently, for the first time, also for single alkane chains\cite{Cui2019nature}).
Our results serve to test our calculation against previous calculation 
and most importantly against recent experimental results 
as well as Landauer based harmonic quantum calculations, 
and demonstrate the applicability robustness of these calculations. 
Furthermore, we present heat conduction results also for a series of conjugated carbon chains 
–- better candidates for molecular electronic transport applications\cite{Crljen2007prl,Garner2018jpcc}
but, as be find, similar to their saturated counterparts in their heat transport behavior.

Section 2 provides details on our simulation technique and our code. 
Section 3 discuss the results of heat conduction properties 
of different types of hydrocarbon chains within molecualr junctions using such approach, 
and compare them to the existing theoreical and experimental data. 
In Section 4 we conclude and give future directions of research in this series.

\section{Model and Calculations}

While the energy spectrum of molecular vibrations encompass a relatively large (~0 – 0.5 eV) frequency range, 
high-frequency vibrations tend to be spatially localized 
and energetically above the cutoff frequency of many solid-state substrates. 
For these reasons, and also because such modes are not populated at room temperature, 
they contribute little to molecular heat transport at that temperature\cite{Segal2003}. 
Molecular heat transport is therefore dominated by lower frequency vibrations, 
for which classical dynamics provide a reasonable approximation. 
Molecular Force-Fields (FF) allow efficient representation of a classical, 
anharmonic molecular Potential Energy Surface (PES) which is the input to the SNEMD studies described below.

\begin{figure}[H]
    \centering
    \includegraphics[width=1\textwidth]{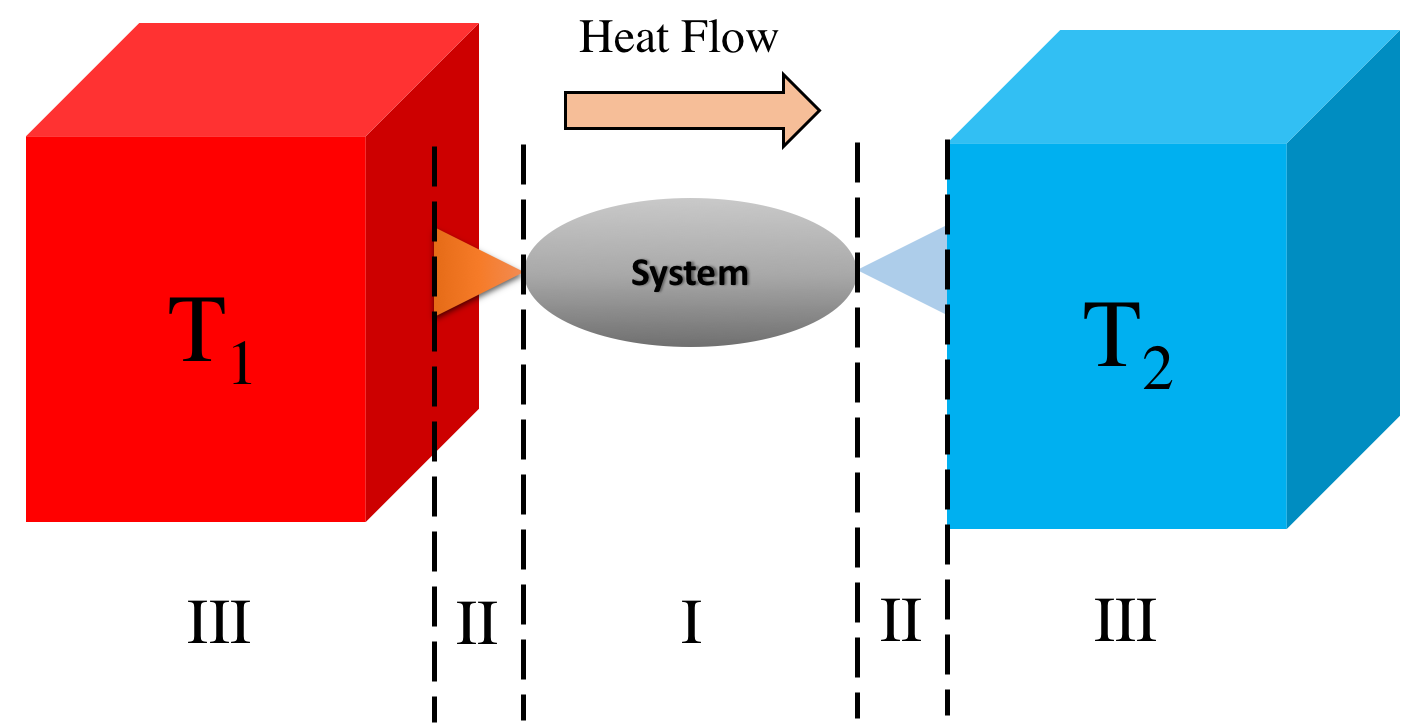}
    \caption{A schematic diagram of the explicit bath model. Region I is the molecular system (including thiol groups); Region II represents the interface and is comprised of explicit layers of metallic materials;
    Region III are implicit baths representing the infinitely large thermal reservoirs, exerting white noise.}
    \label{fig:schematic}
\end{figure}

In the present work we chose to represent the attributes of the thermal environment 
using an explicit thermal bath. 
As shown in figure 2, 
we extend our molecular system with one or more atomic layers of the substrate, 
while the bulk atoms furthest from the molecular system are subjected to Markovian white noise 
which is thus filtered by the explicit substrate layers. 
Specifically, the interfaces between the molecule and the baths (Region III) 
on either side of it are denoted as Region II in the diagram (Figure \ref{fig:schematic}). 
They are modeled using the same Molecular Mechanical Force Fields as the molecular system, 
which are optimized for small organic and organometallic molecules (more details later in this section). 
The interfaces are comprised of an explicit part of the bulk, 
which can be seen as the tips of the measuring apparatus, 
and are usually composed of layers of metallic materials (e.g gold, platinum).

The calculation of the heat current starts 
by representing the potential energy of the whole system 
as a sum of individual interaction terms $V_\tau$, 
where $\tau$ refers to different interaction types, 
for example, the two-body interaction between atom 1 and 2 
or the three-body interaction between atoms 1, 2, and 3. 
We assume an interaction term $V_\tau$ can be further separated 
as a weighted sum over the atoms connected by it, 
weighted according to some partition scheme. 
A generalized analytic formalism has been explored by Torii \textit{et al.}\cite{Torii2008jcp}, explicitly 

\begin{equation}	\label{eqn:potential_partition}
	\begin{split}
		E_{tot}&=\sum_i^{N}\frac{1}{2}m_i\mathbf{v}^2_i+\sum_\tau V_\tau \\V_\tau&=\sum_j^{n(\tau)}U_{\tau,j}, \\ 
	U_{\tau,j}(\{\mathbf{r}_1\ldots\mathbf{r}_{n(\tau)}\})&=C_{\tau,j}V_\tau(\{\mathbf{r}_1\ldots\mathbf{r}_{n(\tau)}\}),\\
	\sum_j^{n(\tau)}C_{\tau,j}&=1,
	\end{split}
\end{equation}

where $n(\tau)$ is the number of atoms connected by the interaction $V_\tau$.
The fraction, $C_{\tau,j}$, of potential energy from $V_\tau$ assigned to atom $j$ is termed as $U_{\tau,j}$.
Assigning specific energies to individual atoms is necessary in order to define atomic energies and energy flows between atoms,
but is obviously somewhat arbitrary. 
In our modeling we chose to assign equal partitioning of each potential energy term between the individual participating atom. 
With such partitioning defined, the heat flux associated with a given atom i in the molecular system, is given by

\begin{equation}
	\label{eqn:heat_flux}
	J_i\equiv\frac{dE_i}{dt}=\frac{d}{dt}\left( \frac{1}{2}m_i\mathbf{v}^2_i+\sum_\tau U_{\tau,i}\right) \\ 
	=\sum_\tau\sum_{j=1}^{n(\tau)} J_{\tau,i j},
\end{equation}
where the heat flux going from atom $j$ to atom $i$
which are connected by $V_\tau$ is defined as,
\begin{equation}
    \label{eqn:heat_flux_tau}
	J_{\tau,i j} = C_{\tau,j}\mathbf{f}_{\tau,i}\cdot \mathbf{v}_i-C_{\tau,i}\mathbf{f}_{\tau,j}\cdot \mathbf{v}_j.
\end{equation}
We have defined $\mathbf{f}_{\tau,j}$ as the force derived from interaction $U_{\tau,j}$.
This is the core expression we use to calculate the inter-atomic heat currents. 
In addition to the inter-atomic force fields we add the effect of the thermal baths through damping forces and random fluctuations from  Langevin Dynamics.
We expect heat current values to plateau as the system tends towards steady-state.

A customized Molecular Dynamics (MD) package built around 
GROningen MAchine for Chemical Simulations
(GROMACS) 4.5\cite{Gromacs4.5} is developed and utilized to conduct the simulations.
The leap-frog algorithm (provided by GROMACS) is used for propagation of the deterministic parts of the system, while Langevin dynamics are used to propagate the stochastic parts of the simulation\cite{Goga2012jctc}.
Unless otherwise stated, the time step is always 1 fs for all runs and the coupling strength between the Markovian bath and outermost layer of explicit bulk (region is 1) is $ps^{-1}$.

First, different utilities, are used to prepare the initial conditions for the simulation. 
These include open source software \textit{Open Babel}\cite{OpenBabel}, \textit{Avogadro}\cite{Avogadro} and other homemade programs and scripts for creating input topologies and indices.
The Universal Force Field (UFF) \cite{UFF} parameters are chosen throughout the simulations.

UFF is one of a few force fields that includes most of the atomic types and bonds across the periodic table, and thus is suitable for organometallic junctions.
As the high frequency carbon-hydrogen bonds often contribute little to the overall vibrational heat conduction, 
it is reasonable to compare side-by-side the effect of with and without hydrogen explicitly appear in the force field. 
This will in principle determine whether it is a good approximation 
to use unified-atom (ua) version of UFF over all-atom(aa) UFF. 
\begin{figure}[H]
    \centering
    \includegraphics[width=1\textwidth]{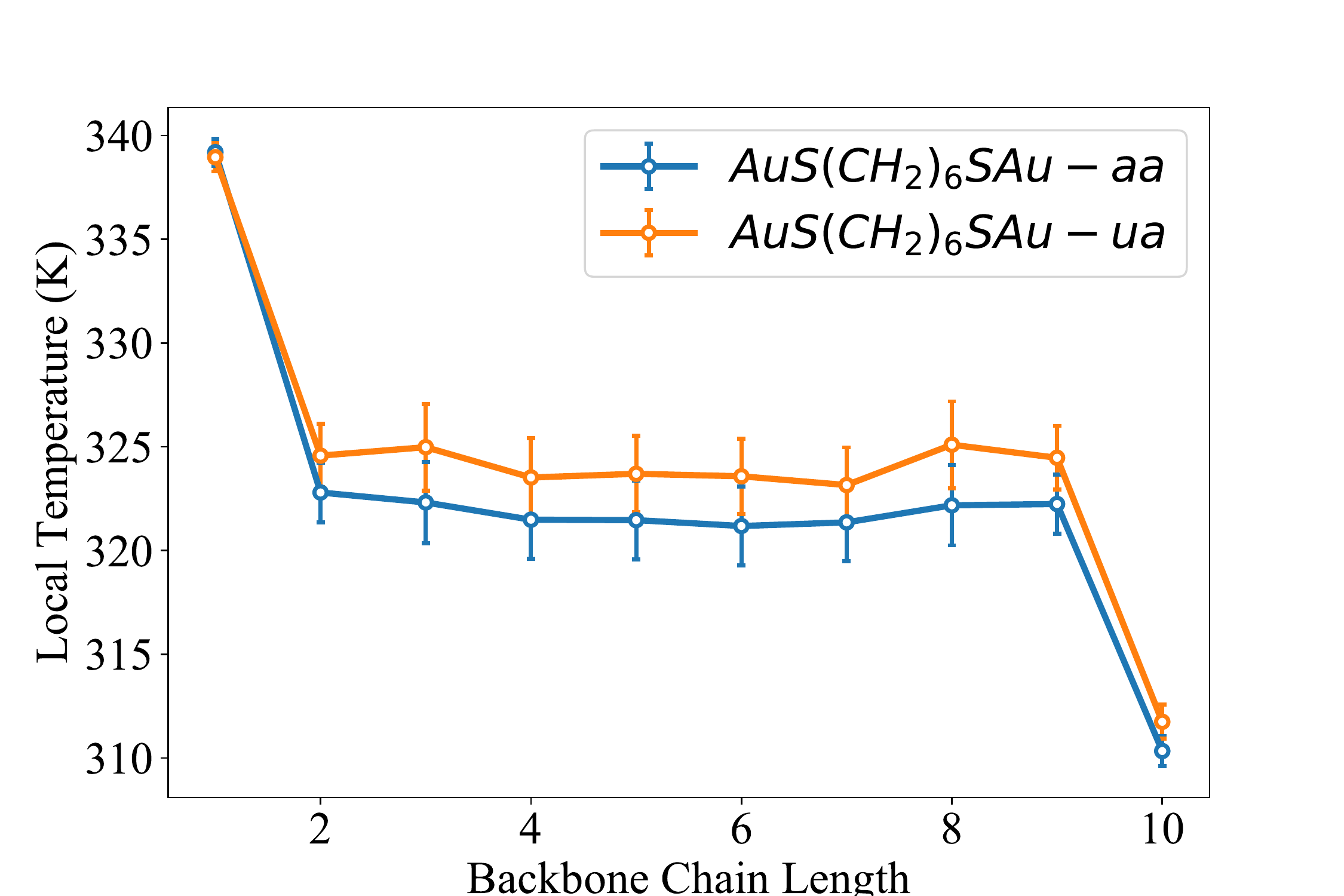}
    \caption{Temperature profile for 1,6-hexanedithiol, comparing UFF all-atom and UFF unified-atom force fields. The horizontal axis is labeled according to the index of the backbone atoms (Sulfur atoms included), and also including the layer of Gold atoms nearest the molecule (the leads), to the left most and right most. The temperatures of the left and right leads are set at 350K and 300K respectively.
    The error bars represent the standard error.(see also Caption in Figure \ref{fig:conductance_layers})}
    \label{fig:T_UFF}
\end{figure}
\begin{table}[ht!]
\centering
\begin{tabular}{||c c c c c||} 
\hline
  & Conductance & SD  & SE. &\\
\hline
 All-Atom & 20.73 & 7.44 & 0.74 &\\ 
 Unified-Atom & 21.77 & 6.40 & 0.64 &\\
 \hline
\end{tabular}
\caption{Heat conductance difference between UFF all-atom and UFF unified-atom force fields, for 1,6-hexanedithiol with three layers of Gold electrodes on each side. 
SD stands for standard deviation, and SE for standard error. All values are given in units of pico-Watts per Kelvin.}
\label{table:UFF}
\end{table}

The local temperatures of the backbone carbons (together with sulfurs and first layer of gold)
are reasonably unchanged when change the UFF-aa to UFF-ua (\ref{fig:T_UFF}).
Table \ref{table:UFF} compares the effect of force-field choice between all-atom and a unified-atom approximation of UFF, on the heat conduction of the hexanedithiold molecule. The relative symmetric difference\footnote{The relative symmetric difference is $\delta(x,y) = \frac{|x-y|}{(x+y)/2}$} between the calculation results for the two force-fields is 4.89\%, and Welch's t-test\footnote{Welch's t-test is $\eta(x,y)=\frac{|\mathbf{E}[x]-\mathbf{E}[y]|}{\sqrt{\sigma_x^2+\sigma_y^2}}$, where $\sigma_x$ is the Standard Deviation in random variable $x$} is 13\%. Therefore, we conclude that the unified-atom approximation is acceptable for our purposes.

The simulation begins by preparing the desired molecular state 
through building of the junction structure, 
and optimizing its geometry to be at the configuration of minimal energy. 
The structure is equilibrated to the average temperature of the baths, 
and then propagated under the boundary conditions of the required temperature bias (e.g. 300K and 350K) until it reaches steady state (usually about a few nanosecond). 
The steady state trajectories are sampled under this specific temperature bias. Pairwise forces between atoms, and between each bath and the atoms coupled to it, are also sampled.

Finally, the heat currents are calculated from the trajectories and forces. The heat currents are then time-averaged.
Ensemble-averages are performed to obtain statistically sound final currents and conductance.

As for Landauer-type calculations, a detailed description of the formalism is provided in the Supporting Information (SI), together with other relevant data and figures.

The heat current equations and computational apparatus described above 
were used to calculate individual heat currents between any two atomic pair within the molecular systems. 
This is not limited to nearest-neighbour bonded atoms (bond stretching interactions), 
but also applies to atoms that are three or four sites apart 
but still interconnected by other interactions parameterized in the force fields (e.g., angle bending, torsion, etc.) 
The heat current flowing from one heat bath to the other bath in the molecular junction, 
can be measured by setting up an imaginary plane 
which is perpendicular to the longitudinal axis of the molecule 
and sum over all the inter-atomic heat currents 
going from one side of the plane to the other side of the plane. 
In steady-state, the heat current through the molecule will be measured the same, 
regardless of where we chose to draw this imaginary plane. 
For computational simplicity, we chose to draw it between region I in figure~\ref{fig:schematic} (molecular) and region II (the substrate interface).
The average thermal conductance is defined as the ratio between this quantity and the temperature bias between the hot and cold baths,
\begin{equation}
    \kappa=\frac{J_{tot}}{T_\text{hot}-T_\text{cold}}.
\end{equation}
In addition to heat fluxes, 
the local temperature of each atom in the conducting molecule 
is calculated from the statistically averaged kinetic energy of the atoms. 
More details regarding our methodology are given at the end of this article and in the Supporting Information (SI).


\section{Results and discussion}

\begin{figure}[H]
    \centering
    \includegraphics[width=1\textwidth]{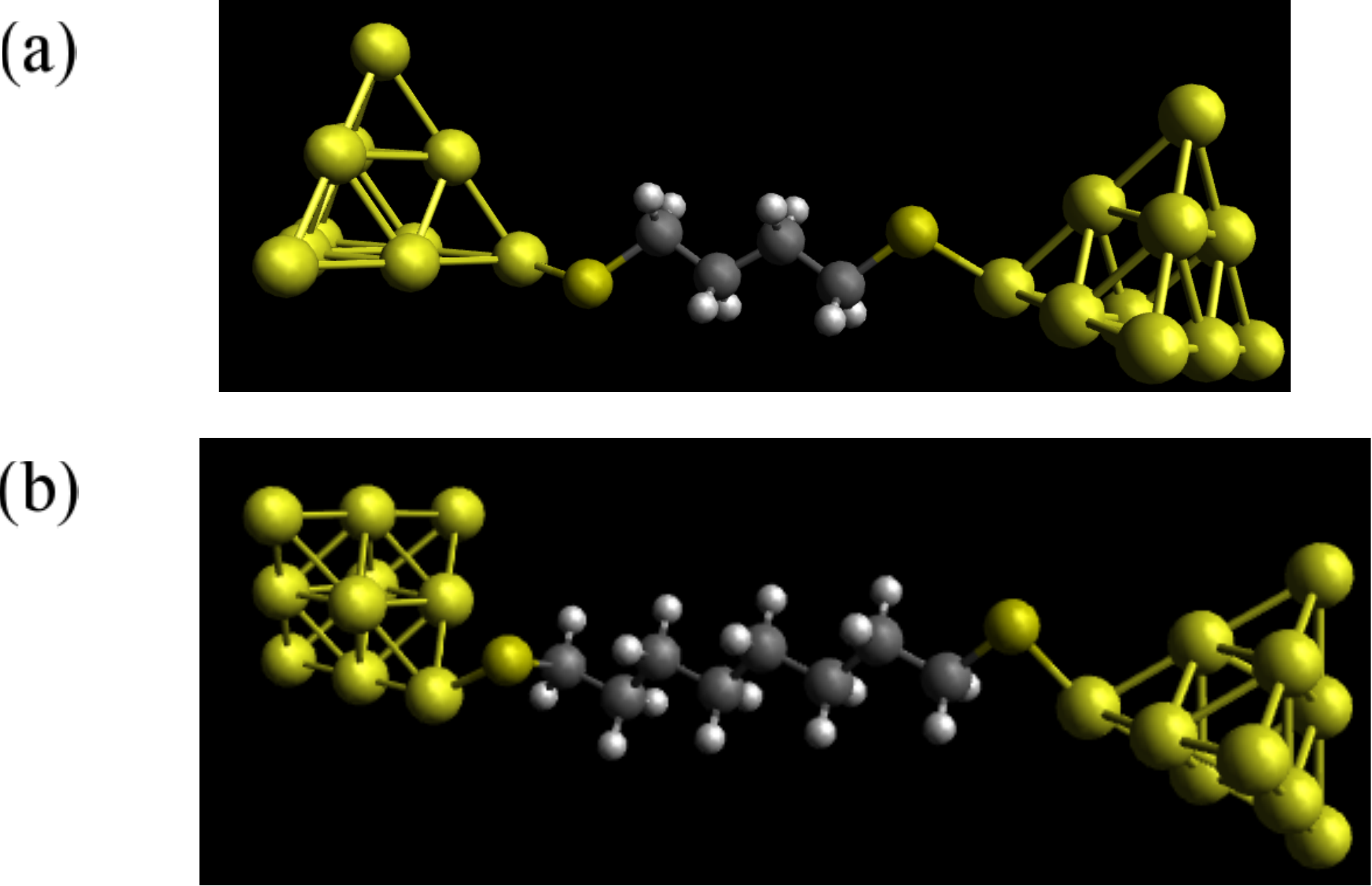}
    \caption{Illustration of some of the alkane molecules studied in the simulation, with three layers of explicit gold bulk. The white noise thermostats are only attached to the layer of gold atoms furthest from the alkane bridge.
    (a) 1,4-butanedithiol(\ch{HS(CH_2)_4SH}); (b) 1,8-octanedithiol(\ch{HS(CH_2)_8SH}).}
    \label{fig:alkanesB}
\end{figure}

The system under investigation is a Single Molecule Junction 
comprising an alkanedithiol \ch{HS(CH_2)_nSH} 
as a molecular bridge connecting several layers of explicit bulk atoms. 
We compared alkanedithiols of various lengths 
(measured by the number of Carbon atoms in the alkane backbone), 
and explicit bulk comprised of one to four layers of gold atoms 
which are further connected to bulk substrates. 
Some examples are illustrated in Figure \ref{fig:alkanesB}. 
For each molecular species, 
we performed MD simulations of the non-equilibrium molecular junction, 
evaluating its steady-state heat transport behavior following the procedure described in Section 2.

For each molecular species, we performed MD simulations of the molecular junction at Nonequilibrium Steady-State (NESS). More computational details are given in the SI.

\begin{figure}[H]
    \centering
\includegraphics[width=1\textwidth]{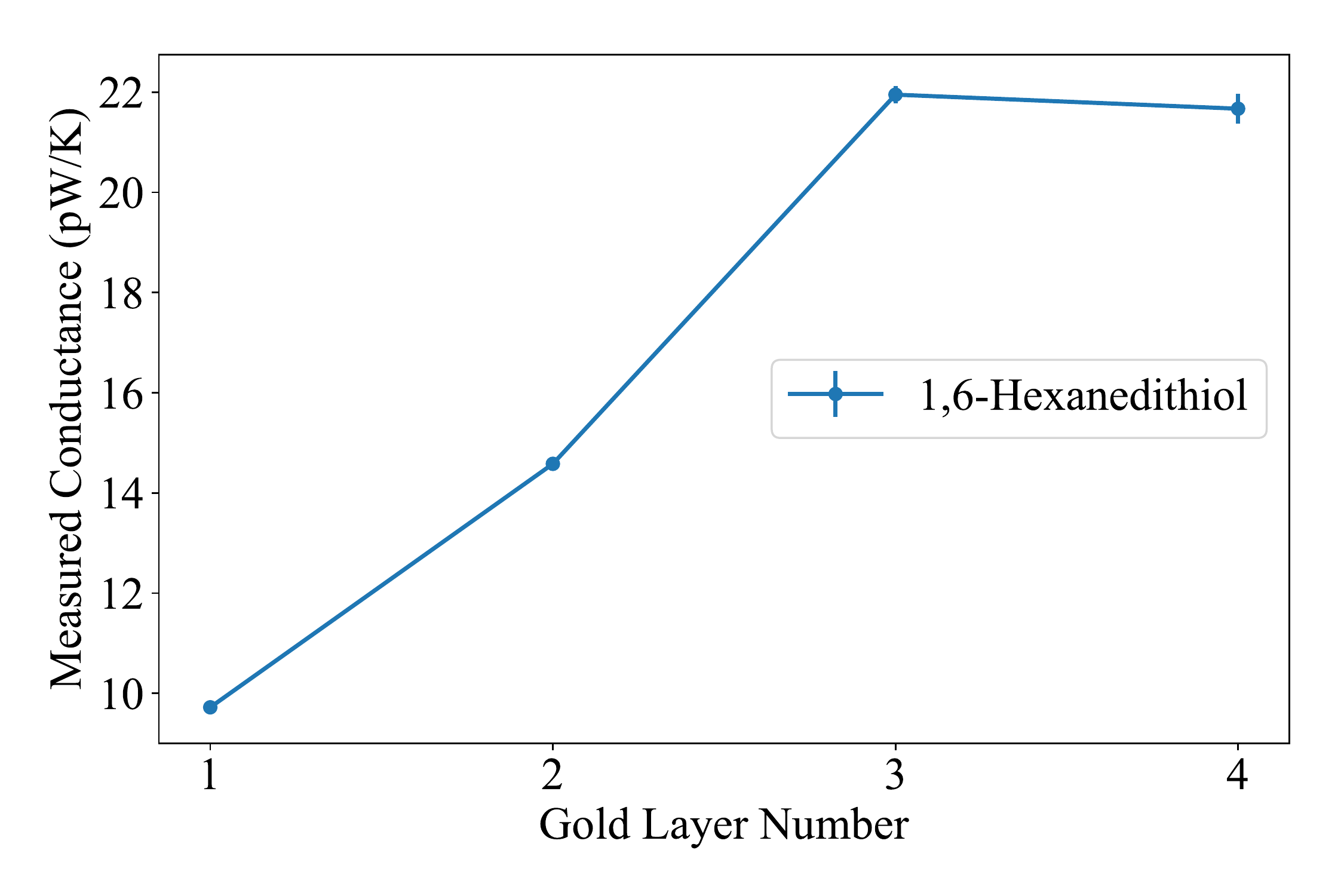} 
    \caption{Heat conductance for the molecule \ch{HS(CH_2)_6SH} given different numbers of gold layers as the explicit bulk.
    The temperature bias is set at 300K to 350K. The bars shown in the figure are the standard errors (SE) 
    of the conductance measurements.
    (SE=Standard Deviation (SD) / square root of the sample size, 
    is a statistical uncertainty indicator of the estimated mean value of the conducted measurements.\cite{SE})
    }
    \label{fig:conductance_layers}
\end{figure}

\begin{figure}[H]
    \centering
    \includegraphics[width=1\textwidth]{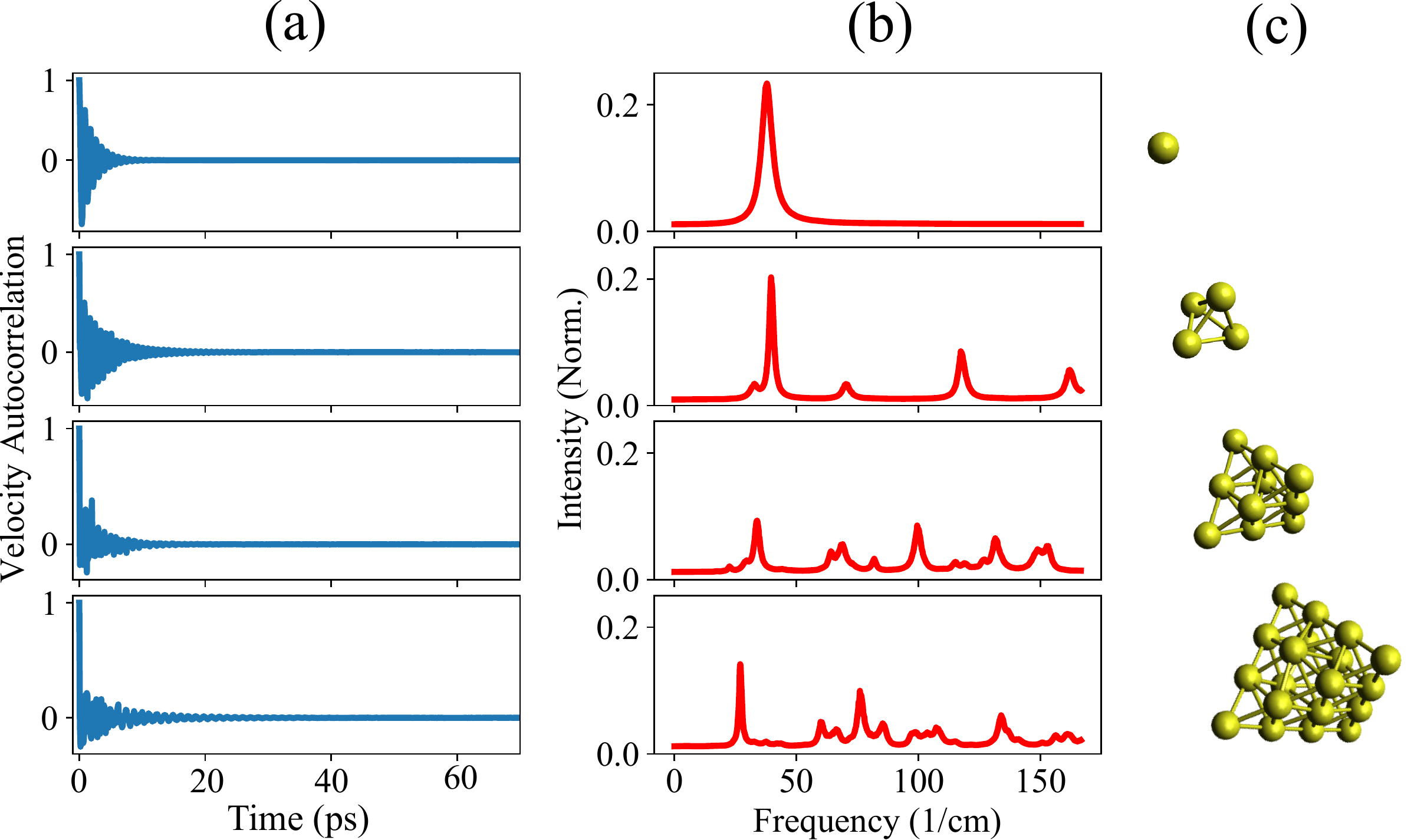}
    \caption{Velocity-velocity autocorrelation functions of the only atom in the first layer of each of four different gold clusters.
    The outer-most layers are attached to Markovian thermal reservoirs at temperature of 300K and the clusters are allowed sufficient time (e.g. a few nanoseconds) to relax to the temperature of the bath.
    Column (a): Velocity time-autocorrelations, $C_{vv}(t)$, which are normalized to the value at $t=t_1-t_2=0$;
    Column (b): The Fourier transforms of the corresponding correlations, normalized across the whole spectra;
    Column (c): Artistic representation of the corresponding gold clusters}
    \label{fig:auto}
\end{figure}

In order to ascertain the relevance of the explicit baths modelling, 
we compared results for molecular heat conductance 
using different numbers of explicit gold layers to represent the bulk. 
Specific simulations are performed on hexanedithiol (Figure \ref{fig:conductance_layers}). 
The calculated conductance appears to converge 
when three layers of explicit gold are used in the substrate representation. 
A similar saturation of layer effect was found also in the other hydrocarbon molecules 
(See SI for results of more alkanedithiols). 
In agreement with this observation, a study by Zhang \textit{et. al.} 
on self-assembled monolayers showed 
that the effect of the baths on the molecular system 
is mainly due to the first few layers of gold substrate\cite{Zhang2010pccp}. 
As seen in Fig. \ref{fig:auto}, 
the observed convergence reflects the convergence in spectral density properties 
of the gold clusters used to represent the thermal baths. 
It should be noted that to enforce the junction geometry, 
the explicit bulk is position-restrained 
by a harmonic force acting on the layer of atoms furthest from the molecule. 
The spectral densities in Fig. \ref{fig:auto} are calculated for the position-restrained clusters 
since the position restraint force is part of the spectral density affecting the molecule. 
The signature of the harmonic position-restraining force is evident at its frequency of about 40 wavenumbers. 
The rest of the spectrum shows frequencies in the range of tens to a few hundred wavenumbers, 
which agrees with experimental measurements of vibrational DOS for gold nanoparticles\cite{Carles2016scirep}.

\begin{figure}[H]
    \centering
    \includegraphics[width=1\textwidth]{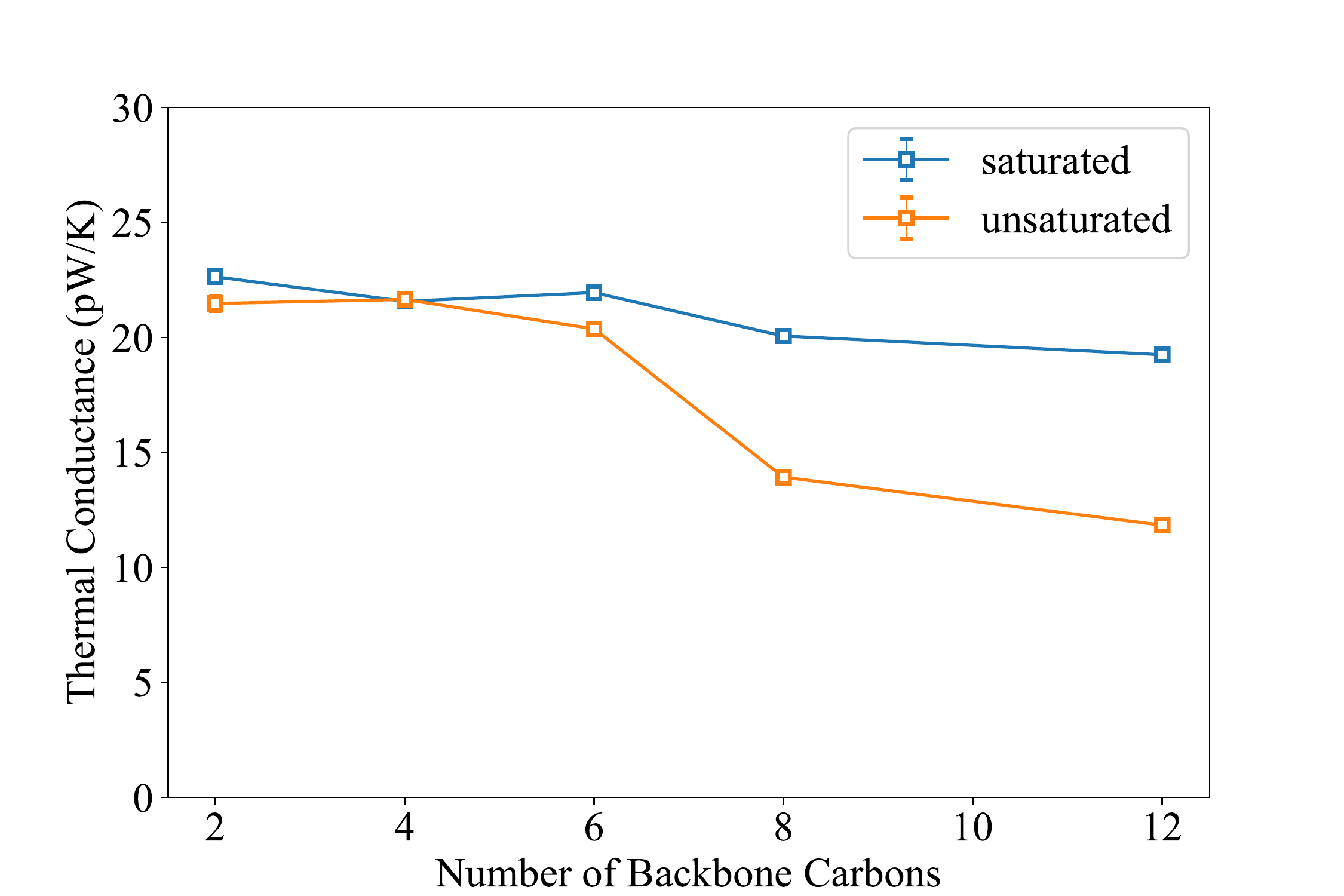}
    \caption{Length-dependent heat conductance of alkane chains (saturated) and triple-bond conjugated hydrocarbon chain (unsaturated) molecules for a temperature bias of 300K and 350K. 
    The error bars represent the SE (see Caption in figure \ref{fig:conductance_layers}) of the conductance measurements.}
    \label{fig:conductance_conjugated}
\end{figure}

\begin{figure}[H]
    \centering
    \includegraphics[width=1\textwidth]{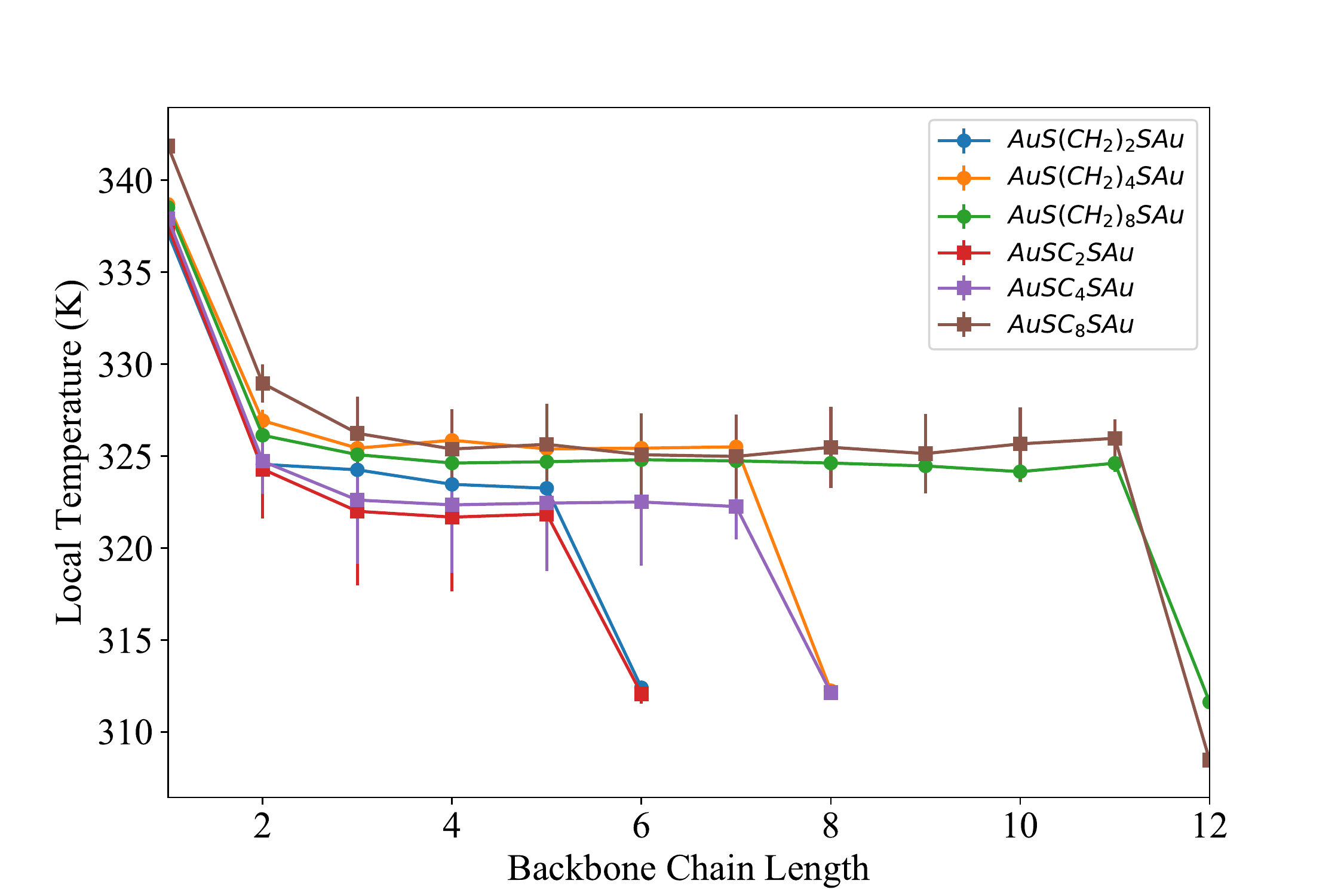}
    \caption{Temperature profile for non-branching alkanedithols of various lengths. In the legend the ones with and without hydrogen atoms are alkanes and conjugated polyyenes respectively. The horizontal axis is labeled according to the index of the backbone atoms (Sulfur atoms included), and also including the layer of gold atoms nearest the molecule (the leads), to the left most and right most. The temperatures of the left and right leads are set at 350K and 300K respectively. That is, atom zero is always the gold atom to the left of the alkanedithiol, atom one is the left Sulfur atom, and the same on the right.
    The error bars represent the SE (see Caption in figure \ref{fig:conductance_layers}) the temperature measurements.}
    \label{fig:T_alkane}
\end{figure}

To study the effect of the molecular structure on the junction conductance, we compared molecules of various lengths and of different degrees of saturation.
Namely, we compared molecules with an alkane backbone (saturated) against molecules with a conjugated polyyne backbone (unsaturated)\footnote{A polyyne is an organic compound with alternating single and triple bonds;
that is, a series of consecutive alkynes, $\left(- C \equiv C - \right)_n$ with $n$ greater than 1.}.
Unless otherwise specified, 
the simulation results displayed below were obtained 
using three explicit atomic layers for the gold substrates. 
Figure~\ref{fig:conductance_conjugated} shows 
that heat conductance of shorter molecules ($n < 6$) 
is not strongly affected by carbon bond saturation, 
however longer unsaturated chains exhibit lower conductance than their saturated counterparts. 
This observation stands in contrast to the higher electronic transport properties of conjugated chain molecules\cite{Crljen2007prl}, 
and may be explained by the difference in current carriers of thermal and electronic transport in these systems: 
While the delocalized electrons in conjugated molecules may contribute much to the overall electronic conduction, 
heat conduction, which is dominated by phonon transport, 
is mostly determined by bond structure and vibrational modes in the molecular system.
The steady state temperature profiles associated with 
the results of Figure \ref{fig:conductance_conjugated} are displayed in Figure \ref{fig:T_alkane}. 
The bias (300K - 350K) clearly exceeds the regime of validity of linear response, 
yet is more realistic with respect to existing experimental setups
\cite{Meier2014prl,Wang2006apl,Majumdar2015nl,Cui2019nature}. 

The temperature profiles show that most of the thermal resistance is interfacial. 
Even the longer molecules are homogeneous enough 
that the temperature profile does not slope significantly, 
which could point to a ballistic regime of heat conduction in the molecular bridge. 
This can be ascribed to the explicit modelling of the molecule-bulk interface at the atomic level.
The features of interfacial temperature already show some filtering effects from the white baths, 
noting the first layer of gold bulk on the left is about 10 degrees lower than external hot reservoir 
and right first layer 10 degrees higher than the external cold reservoir.

\begin{figure}
    \centering
    \includegraphics[width=1\textwidth]{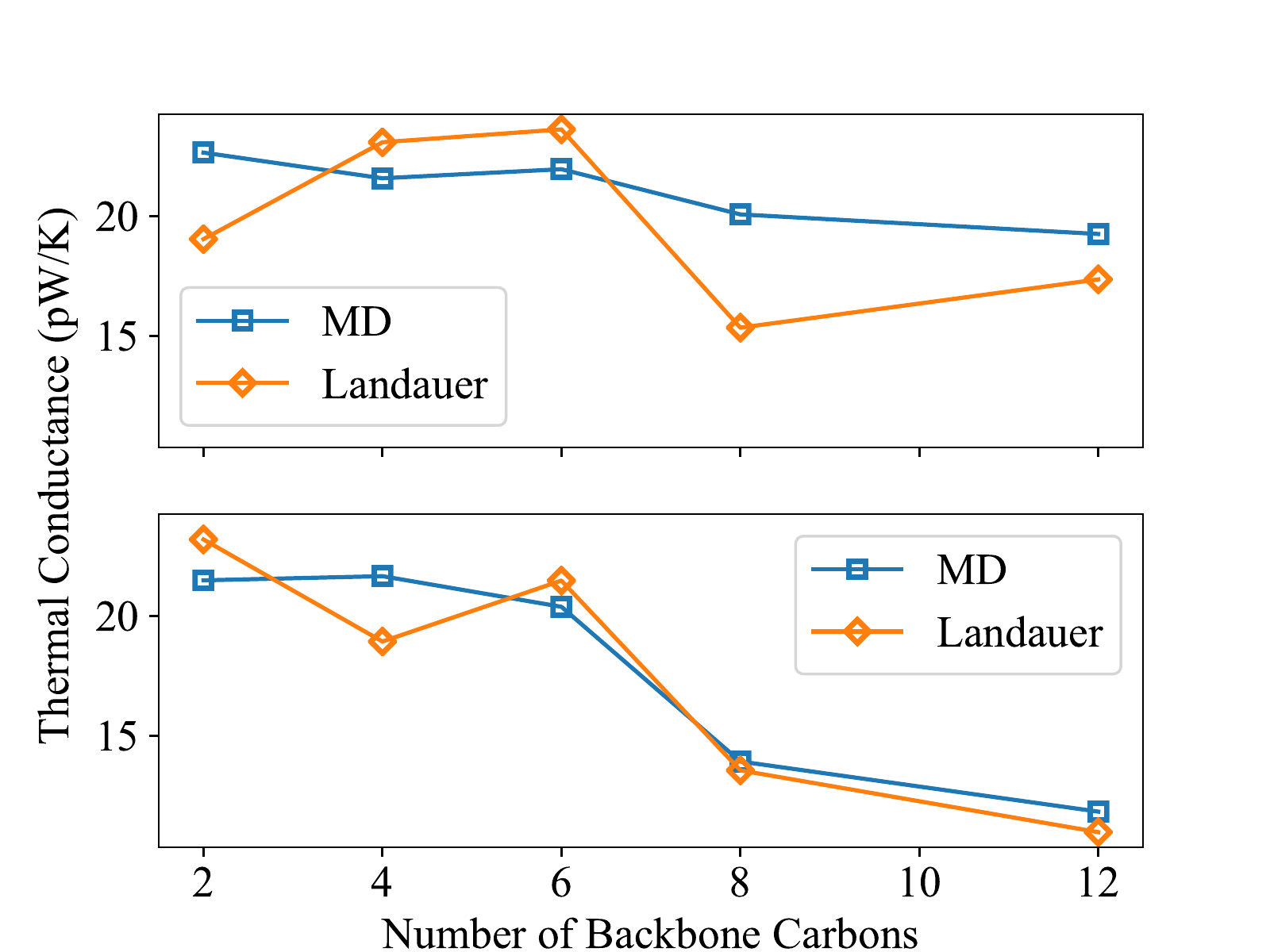}
    \caption{Length-dependent heat conductance of alkane chains (saturated, in the upper panel) and triple-bond conjugated hydrocarbon chain (unsaturated, in the lower panel) molecules, obtained with GROMACS MD simulations and Landauer-type quantum calculations, respectively. In all cases, thiol head-groups are attached, and the currents are calculated across a temperature bias from 350K to 300K, with conductance defined as the ratio between heat current and temperature bias (50K).}
    \label{fig:conductance_MD_vs_Landauer}
\end{figure}

Next, consider the results obtained from the quantum-mechanical harmonic model calculation. 
The theoretical and numerical details of the Landauer approach for calculating heat conductance 
has been specified in the Support Information. 
Figure \ref{fig:conductance_MD_vs_Landauer} compares the results of this model 
to those obtained from the classical MD simulation. 
While a certain degree of divergence occurs of these two methods
in the results of alkanedithiolds 
(which we will explain in the later parts in the section), 
the remarkable agreement in polyynes molecules 
indicates that heat conduction in these systems 
is determined by the harmonic part of the force field 
and dominated by modes from the lower frequency range of the molecular spectrum. 

Finally, consider the absolute values of the calculatioted heat conductions. 
In general, our results are in agreement with the most recent experimental measurements
for heat conduction in SMJs \cite{Cui2019nature}. 
For the saturated alkanes, the Landauer conductance starts to fluctuate and then goes down as the chain grows even longer (See SI for more data), 
while the MD results decrease slowly and stay relatively stable around the value 20 pW/K. 
Though the non-monotonic behavior of the Landauer calculations align with an independent ab initio Landauer-type calculation\cite{Kloeckner2016},
MD simulations align more closely with the experimental data and trend\cite{Cui2019nature},
indicating anharmonicity plays a tunning role in molecular thermal transport.
It is interesting and intuitive to see heat conduction from a normal mode harmonic perspective
(i.e. the normal mode density, localization, and transmission probablity, see SI for detailed analysis),
but at room temperature classical MD seems do better in taking into account both harmonic and anharmonic effects
by utlizing full-force-field interactions of the molecular potential energy.


\section{Conclusion}
In summary, we have presented results of classical MD simulations using stochastic Langevin thermal baths, 
as well as results of a quantum calculation based on the harmonic part of the molecular force field, 
for the steady-state heat conduction of molecular junctions comprising saturated and conjugated hydrocarbon chains connecting gold leads. 
The multiple layers of explict gold substrates act as filters of larger environmental white noise
and bring characteristic bath effects to the heat conducting molecular systems under investigation.
The high degree of agreement between our simulations and the most recent experimental measurements\cite{Cui2019nature}
also validates the methods and numerical tools we use.
For the alkanedithiols in particular,
MD simulation agrees better with the experiment than Landauer-type quantum calculation in conditions of room temperature and with large bias. 
This may hint that at ambient conditions, 
the explicit treatment of quantum effects is less relevant than explicit treatment of anharmonicity, 
and linear response less relevant than the behavior of systems far-from-equilibrium. 
For conjugated hydrocarbons,
while previous studies have shown unusually high electronic conduction of polyynes\cite{Crljen2007prl}, 
our study shows that they have a lower thermal conductance than their saturated counterparts. 
which  might indicate they are potentially good candidates for thermoelectric nanomaterials.
Still the effects of complex molecular structures and topologies on nanoscale heat conduction are still not clear,
and will be future directions of our research in the group by making full use of the atomistic-resolution afforded by our approach.


\begin{acknowledgement}
The research of AN is supported by the Israel-U. S. Binational Science Foundation, 
the German Research Foundation (DFG TH 820/11-1), the U. S. National ScienceFoundation(GrantNo. CHE1665291),and the University of Pennsylvania.
\end{acknowledgement}






\bibliography{bib/j,bib/references,bib/achemso-demo}

\providecommand{\latin}[1]{#1}
\makeatletter
\providecommand{\doi}
  {\begingroup\let\do\@makeother\dospecials
  \catcode`\{=1 \catcode`\}=2 \doi@aux}
\providecommand{\doi@aux}[1]{\endgroup\texttt{#1}}
\makeatother
\providecommand*\mcitethebibliography{\thebibliography}
\csname @ifundefined\endcsname{endmcitethebibliography}
  {\let\endmcitethebibliography\endthebibliography}{}
\begin{mcitethebibliography}{48}
\providecommand*\natexlab[1]{#1}
\providecommand*\mciteSetBstSublistMode[1]{}
\providecommand*\mciteSetBstMaxWidthForm[2]{}
\providecommand*\mciteBstWouldAddEndPuncttrue
  {\def\EndOfBibitem{\unskip.}}
\providecommand*\mciteBstWouldAddEndPunctfalse
  {\let\EndOfBibitem\relax}
\providecommand*\mciteSetBstMidEndSepPunct[3]{}
\providecommand*\mciteSetBstSublistLabelBeginEnd[3]{}
\providecommand*\EndOfBibitem{}
\mciteSetBstSublistMode{f}
\mciteSetBstMaxWidthForm{subitem}{(\alph{mcitesubitemcount})}
\mciteSetBstSublistLabelBeginEnd
  {\mcitemaxwidthsubitemform\space}
  {\relax}
  {\relax}

\bibitem[Segal and Agarwalla(2016)Segal, and Agarwalla]{Segal2016arpc}
Segal,~D.; Agarwalla,~B.~K. Vibrational Heat Transport in Molecular Junctions.
  \emph{Annu. Rev. Phys. Chem.} \textbf{2016}, \emph{67}, 185--209\relax
\mciteBstWouldAddEndPuncttrue
\mciteSetBstMidEndSepPunct{\mcitedefaultmidpunct}
{\mcitedefaultendpunct}{\mcitedefaultseppunct}\relax
\EndOfBibitem
\bibitem[Pop(2010)]{Pop2010}
Pop,~E. Energy dissipation and transport in nanoscale devices. \emph{Nano Res.}
  \textbf{2010}, \emph{3}, 147--169\relax
\mciteBstWouldAddEndPuncttrue
\mciteSetBstMidEndSepPunct{\mcitedefaultmidpunct}
{\mcitedefaultendpunct}{\mcitedefaultseppunct}\relax
\EndOfBibitem
\bibitem[Li \latin{et~al.}(2012)Li, Ren, Wang, Zhang, H\"anggi, and
  Li]{Li2012rmp}
Li,~N.; Ren,~J.; Wang,~L.; Zhang,~G.; H\"anggi,~P.; Li,~B. \textit{Colloquium}
  : Phononics: Manipulating heat flow with electronic analogs and beyond.
  \emph{Rev. Mod. Phys.} \textbf{2012}, \emph{84}, 1045--1066\relax
\mciteBstWouldAddEndPuncttrue
\mciteSetBstMidEndSepPunct{\mcitedefaultmidpunct}
{\mcitedefaultendpunct}{\mcitedefaultseppunct}\relax
\EndOfBibitem
\bibitem[Dubi and Di~Ventra(2011)Dubi, and Di~Ventra]{Dubi2011}
Dubi,~Y.; Di~Ventra,~M. \textit{Colloquium} : Heat flow and thermoelectricity
  in atomic and molecular junctions. \emph{Rev. Mod. Phys.} \textbf{2011},
  \emph{83}, 131--155\relax
\mciteBstWouldAddEndPuncttrue
\mciteSetBstMidEndSepPunct{\mcitedefaultmidpunct}
{\mcitedefaultendpunct}{\mcitedefaultseppunct}\relax
\EndOfBibitem
\bibitem[Yamamoto and Watanabe(2006)Yamamoto, and Watanabe]{Yamamoto2006}
Yamamoto,~T.; Watanabe,~K. Nonequilibrium Green's Function Approach to Phonon
  Transport in Defective Carbon Nanotubes. \emph{Phys. Rev. Lett.}
  \textbf{2006}, \emph{96}, 255503\relax
\mciteBstWouldAddEndPuncttrue
\mciteSetBstMidEndSepPunct{\mcitedefaultmidpunct}
{\mcitedefaultendpunct}{\mcitedefaultseppunct}\relax
\EndOfBibitem
\bibitem[Wang \latin{et~al.}(2006)Wang, Wang, and Zeng]{Wang2006prb}
Wang,~J.-S.; Wang,~J.; Zeng,~N. Nonequilibrium Green's function approach to
  mesoscopic thermal transport. \emph{Phys. Rev. B} \textbf{2006}, \emph{74},
  033408\relax
\mciteBstWouldAddEndPuncttrue
\mciteSetBstMidEndSepPunct{\mcitedefaultmidpunct}
{\mcitedefaultendpunct}{\mcitedefaultseppunct}\relax
\EndOfBibitem
\bibitem[Landauer(1957)]{landauer1957}
Landauer,~R. Spatial {Variation} of {Currents} and {Fields} {Due} to
  {Localized} {Scatterers} in {Metallic} {Conduction}. \emph{IBM Journal of
  Research and Development} \textbf{1957}, \emph{1}, 223\relax
\mciteBstWouldAddEndPuncttrue
\mciteSetBstMidEndSepPunct{\mcitedefaultmidpunct}
{\mcitedefaultendpunct}{\mcitedefaultseppunct}\relax
\EndOfBibitem
\bibitem[Zhang \latin{et~al.}(2005)Zhang, Lussetti, de~Souza, and
  Müller-Plathe]{Zhang2005jpcb}
Zhang,~M.; Lussetti,~E.; de~Souza,~L. E.~S.; Müller-Plathe,~F. Thermal
  Conductivities of Molecular Liquids by Reverse Nonequilibrium Molecular
  Dynamics. \emph{J. Phys. Chem. B} \textbf{2005}, \emph{109},
  15060--15067\relax
\mciteBstWouldAddEndPuncttrue
\mciteSetBstMidEndSepPunct{\mcitedefaultmidpunct}
{\mcitedefaultendpunct}{\mcitedefaultseppunct}\relax
\EndOfBibitem
\bibitem[Lukes \latin{et~al.}(2000)Lukes, Li, Liang, and Tien]{Lukes2000jht}
Lukes,~J.~R.; Li,~D.~Y.; Liang,~X.-G.; Tien,~C.-L. {Molecular Dynamics Study of
  Solid Thin-Film Thermal Conductivity}. \emph{J. Heat Transfer} \textbf{2000},
  \emph{122}, 536--543\relax
\mciteBstWouldAddEndPuncttrue
\mciteSetBstMidEndSepPunct{\mcitedefaultmidpunct}
{\mcitedefaultendpunct}{\mcitedefaultseppunct}\relax
\EndOfBibitem
\bibitem[Berber \latin{et~al.}(2000)Berber, Kwon, and
  Tom{\'{a}}nek]{Berber2000prl}
Berber,~S.; Kwon,~Y.-k.; Tom{\'{a}}nek,~D. {Unusually high thermal conductivity
  of carbon nanotubes}. \emph{Phys. Rev. Lett.} \textbf{2000}, \emph{84},
  4613--4616\relax
\mciteBstWouldAddEndPuncttrue
\mciteSetBstMidEndSepPunct{\mcitedefaultmidpunct}
{\mcitedefaultendpunct}{\mcitedefaultseppunct}\relax
\EndOfBibitem
\bibitem[Ness \latin{et~al.}(2017)Ness, Stella, Lorenz, and
  Kantorovich]{Ness2017jcp}
Ness,~H.; Stella,~L.; Lorenz,~C.~D.; Kantorovich,~L. Nonequilibrium generalised
  Langevin equation for the calculation of heat transport properties in model
  1D atomic chains coupled to two 3D thermal baths. \emph{J. Chem. Phys.}
  \textbf{2017}, \emph{146}, 164103\relax
\mciteBstWouldAddEndPuncttrue
\mciteSetBstMidEndSepPunct{\mcitedefaultmidpunct}
{\mcitedefaultendpunct}{\mcitedefaultseppunct}\relax
\EndOfBibitem
\bibitem[Ladd \latin{et~al.}(1986)Ladd, Moran, and Hoover]{Ladd1986prb}
Ladd,~A. J.~C.; Moran,~B.; Hoover,~W.~G. Lattice thermal conductivity: A
  comparison of molecular dynamics and anharmonic lattice dynamics. \emph{Phys.
  Rev. B} \textbf{1986}, \emph{34}, 5058--5064\relax
\mciteBstWouldAddEndPuncttrue
\mciteSetBstMidEndSepPunct{\mcitedefaultmidpunct}
{\mcitedefaultendpunct}{\mcitedefaultseppunct}\relax
\EndOfBibitem
\bibitem[Volz and Chen(2000)Volz, and Chen]{Volz2000prb}
Volz,~S.~G.; Chen,~G. {Molecular-dynamics simulation of thermal conductivity of
  silicon crystals}. \emph{Phys. Rev. B} \textbf{2000}, \emph{61},
  2651--2656\relax
\mciteBstWouldAddEndPuncttrue
\mciteSetBstMidEndSepPunct{\mcitedefaultmidpunct}
{\mcitedefaultendpunct}{\mcitedefaultseppunct}\relax
\EndOfBibitem
\bibitem[Che \latin{et~al.}(2000)Che, Çağın, Deng, and Goddard]{Che2000jcp}
Che,~J.; Çağın,~T.; Deng,~W.; Goddard,~W.~A. Thermal conductivity of diamond
  and related materials from molecular dynamics simulations. \emph{J. Chem.
  Phys.} \textbf{2000}, \emph{113}, 6888--6900\relax
\mciteBstWouldAddEndPuncttrue
\mciteSetBstMidEndSepPunct{\mcitedefaultmidpunct}
{\mcitedefaultendpunct}{\mcitedefaultseppunct}\relax
\EndOfBibitem
\bibitem[Schelling \latin{et~al.}(2002)Schelling, Phillpot, and
  Keblinski]{Schelling2002prb}
Schelling,~P.~K.; Phillpot,~S.~R.; Keblinski,~P. {Comparison of atomic-level
  simulation methods for computing thermal conductivity}. \emph{Physical Review
  B - Condensed Matter and Materials Physics} \textbf{2002}, \emph{65},
  1--12\relax
\mciteBstWouldAddEndPuncttrue
\mciteSetBstMidEndSepPunct{\mcitedefaultmidpunct}
{\mcitedefaultendpunct}{\mcitedefaultseppunct}\relax
\EndOfBibitem
\bibitem[McGaughey and Kaviany(2004)McGaughey, and
  Kaviany]{McGaughey2004ijhmt1}
McGaughey,~A.; Kaviany,~M. Thermal conductivity decomposition and analysis
  using molecular dynamics simulations. Part I. Lennard-Jones argon. \emph{Int.
  J. Heat Mass Transf.} \textbf{2004}, \emph{47}, 1783 -- 1798\relax
\mciteBstWouldAddEndPuncttrue
\mciteSetBstMidEndSepPunct{\mcitedefaultmidpunct}
{\mcitedefaultendpunct}{\mcitedefaultseppunct}\relax
\EndOfBibitem
\bibitem[McGaughey and Kaviany(2004)McGaughey, and
  Kaviany]{McGaughey2004ijhmt2}
McGaughey,~A.; Kaviany,~M. Thermal conductivity decomposition and analysis
  using molecular dynamics simulations: Part II. Complex silica structures.
  \emph{Int. J. Heat Mass Transf.} \textbf{2004}, \emph{47}, 1799 -- 1816\relax
\mciteBstWouldAddEndPuncttrue
\mciteSetBstMidEndSepPunct{\mcitedefaultmidpunct}
{\mcitedefaultendpunct}{\mcitedefaultseppunct}\relax
\EndOfBibitem
\bibitem[Chen and Kumar(2012)Chen, and Kumar]{Chen2012jap}
Chen,~L.; Kumar,~S. Thermal transport in graphene supported on copper. \emph{J.
  Appl. Phys.} \textbf{2012}, \emph{112}, 043502\relax
\mciteBstWouldAddEndPuncttrue
\mciteSetBstMidEndSepPunct{\mcitedefaultmidpunct}
{\mcitedefaultendpunct}{\mcitedefaultseppunct}\relax
\EndOfBibitem
\bibitem[Sellan \latin{et~al.}(2010)Sellan, Landry, Turney, McGaughey, and
  Amon]{Sellan2010prb}
Sellan,~D.~P.; Landry,~E.~S.; Turney,~J.~E.; McGaughey,~A. J.~H.; Amon,~C.~H.
  {Size effects in molecular dynamics thermal conductivity predictions}.
  \emph{Phys. Rev. B} \textbf{2010}, \emph{81}, 1--10\relax
\mciteBstWouldAddEndPuncttrue
\mciteSetBstMidEndSepPunct{\mcitedefaultmidpunct}
{\mcitedefaultendpunct}{\mcitedefaultseppunct}\relax
\EndOfBibitem
\bibitem[Baranyai(1996)]{Baranyai1996}
Baranyai,~A. {Heat flow studies for large temperature gradients by molecular
  dynamics simulation}. \emph{Phys. Rev. E} \textbf{1996}, \emph{54},
  6911--6917\relax
\mciteBstWouldAddEndPuncttrue
\mciteSetBstMidEndSepPunct{\mcitedefaultmidpunct}
{\mcitedefaultendpunct}{\mcitedefaultseppunct}\relax
\EndOfBibitem
\bibitem[Poetzsch and Bottger(1994)Poetzsch, and Bottger]{Poetzsch1994}
Poetzsch,~R. H.~H.; Bottger,~H. Interplay of disorder and anharmonicity in heat
  conduction: Molecular-dynamics study. \emph{Phys. Rev. B} \textbf{1994},
  \emph{50}, 757--764\relax
\mciteBstWouldAddEndPuncttrue
\mciteSetBstMidEndSepPunct{\mcitedefaultmidpunct}
{\mcitedefaultendpunct}{\mcitedefaultseppunct}\relax
\EndOfBibitem
\bibitem[Oligschleger(1999)]{Oligschleger1999}
Oligschleger,~C. {Simulation of thermal conductivity and heat transport in
  solids}. \emph{Phys. Rev. B} \textbf{1999}, \emph{59}, 4125--4133\relax
\mciteBstWouldAddEndPuncttrue
\mciteSetBstMidEndSepPunct{\mcitedefaultmidpunct}
{\mcitedefaultendpunct}{\mcitedefaultseppunct}\relax
\EndOfBibitem
\bibitem[Jiang \latin{et~al.}(2010)Jiang, Lan, Wang, and Li]{Jiang2010jap}
Jiang,~J.~W.; Lan,~J.; Wang,~J.~S.; Li,~B. {Isotopic effects on the thermal
  conductivity of graphene nanoribbons: Localization mechanism}. \emph{J. Appl.
  Phys.} \textbf{2010}, \emph{107}\relax
\mciteBstWouldAddEndPuncttrue
\mciteSetBstMidEndSepPunct{\mcitedefaultmidpunct}
{\mcitedefaultendpunct}{\mcitedefaultseppunct}\relax
\EndOfBibitem
\bibitem[Jund and Jullien(1999)Jund, and Jullien]{Jund1999}
Jund,~P.; Jullien,~R. {Molecular-dynamics calculation of the thermal
  conductivity of vitreous silica}. \emph{Phys. Rev. B} \textbf{1999},
  \emph{59}, 13707--13711\relax
\mciteBstWouldAddEndPuncttrue
\mciteSetBstMidEndSepPunct{\mcitedefaultmidpunct}
{\mcitedefaultendpunct}{\mcitedefaultseppunct}\relax
\EndOfBibitem
\bibitem[M{\"{u}}ller-Plathe(1997)]{Muller-Plathe1997}
M{\"{u}}ller-Plathe,~F. {A simple nonequilibrium molecular dynamics method for
  calculating the thermal conductivity}. \emph{J. Chem. Phys.} \textbf{1997},
  \emph{106}, 6082--6085\relax
\mciteBstWouldAddEndPuncttrue
\mciteSetBstMidEndSepPunct{\mcitedefaultmidpunct}
{\mcitedefaultendpunct}{\mcitedefaultseppunct}\relax
\EndOfBibitem
\bibitem[M{\"{u}}ller-Plathe and Reith(1999)M{\"{u}}ller-Plathe, and
  Reith]{Muller-Plathe1999}
M{\"{u}}ller-Plathe,~F.; Reith,~D. {Cause and effect reversed in
  non-equilibrium molecular dynamics: An easy route to transport coefficients}.
  \emph{Computational and Theoretical Polymer Science} \textbf{1999}, \emph{9},
  203--209\relax
\mciteBstWouldAddEndPuncttrue
\mciteSetBstMidEndSepPunct{\mcitedefaultmidpunct}
{\mcitedefaultendpunct}{\mcitedefaultseppunct}\relax
\EndOfBibitem
\bibitem[Bagri \latin{et~al.}(2011)Bagri, Kim, Ruoff, and Shenoy]{Bagri2011nl}
Bagri,~A.; Kim,~S.~P.; Ruoff,~R.~S.; Shenoy,~V.~B. {Thermal transport across
  twin grain boundaries in polycrystalline graphene from nonequilibrium
  molecular dynamics simulations}. \emph{Nano Letters} \textbf{2011},
  \emph{11}, 3917--3921\relax
\mciteBstWouldAddEndPuncttrue
\mciteSetBstMidEndSepPunct{\mcitedefaultmidpunct}
{\mcitedefaultendpunct}{\mcitedefaultseppunct}\relax
\EndOfBibitem
\bibitem[Dong \latin{et~al.}(2014)Dong, Wen, and Melnik]{Dong2014scirep}
Dong,~H.; Wen,~B.; Melnik,~R. {Relative importance of grain boundaries and size
  effects in thermal conductivity of nanocrystalline materials}. \emph{Sci.
  Rep.} \textbf{2014}, \emph{4}, 7037\relax
\mciteBstWouldAddEndPuncttrue
\mciteSetBstMidEndSepPunct{\mcitedefaultmidpunct}
{\mcitedefaultendpunct}{\mcitedefaultseppunct}\relax
\EndOfBibitem
\bibitem[Tang and Kulkarni(2013)Tang, and Kulkarni]{Tang2013apl}
Tang,~S.; Kulkarni,~Y. The interplay between strain and size effects on the
  thermal conductance of grain boundaries in graphene. \emph{Appl. Phys. Lett.}
  \textbf{2013}, \emph{103}, 213113\relax
\mciteBstWouldAddEndPuncttrue
\mciteSetBstMidEndSepPunct{\mcitedefaultmidpunct}
{\mcitedefaultendpunct}{\mcitedefaultseppunct}\relax
\EndOfBibitem
\bibitem[Si \latin{et~al.}(2017)Si, Wang, Fan, Feng, and Cao]{Si2017ijhmt}
Si,~C.; Wang,~X.-D.; Fan,~Z.; Feng,~Z.-H.; Cao,~B.-Y. Impacts of potential
  models on calculating the thermal conductivity of graphene using
  non-equilibrium molecular dynamics simulations. \emph{Int. J. Heat Mass
  Transf.} \textbf{2017}, \emph{107}, 450 -- 460\relax
\mciteBstWouldAddEndPuncttrue
\mciteSetBstMidEndSepPunct{\mcitedefaultmidpunct}
{\mcitedefaultendpunct}{\mcitedefaultseppunct}\relax
\EndOfBibitem
\bibitem[Goga \latin{et~al.}(2012)Goga, Rzepiela, de~Vries, Marrink, and
  Berendsen]{Goga2012jctc}
Goga,~N.; Rzepiela,~A.~J.; de~Vries,~A.~H.; Marrink,~S.~J.; Berendsen,~H. J.~C.
  Efficient Algorithms for Langevin and DPD Dynamics. \emph{J. Chem. Theory
  Comput.} \textbf{2012}, \emph{8}, 3637--3649\relax
\mciteBstWouldAddEndPuncttrue
\mciteSetBstMidEndSepPunct{\mcitedefaultmidpunct}
{\mcitedefaultendpunct}{\mcitedefaultseppunct}\relax
\EndOfBibitem
\bibitem[Segal \latin{et~al.}(2003)Segal, Nitzan, and H\"anggi]{Segal2003}
Segal,~D.; Nitzan,~A.; H\"anggi,~P. Thermal conductance through molecular
  wires. \emph{J. Chem. Phys.} \textbf{2003}, \emph{119}, 6840--6855\relax
\mciteBstWouldAddEndPuncttrue
\mciteSetBstMidEndSepPunct{\mcitedefaultmidpunct}
{\mcitedefaultendpunct}{\mcitedefaultseppunct}\relax
\EndOfBibitem
\bibitem[Kl{\"o}ckner \latin{et~al.}(2016)Kl{\"o}ckner, B{\"u}rkle, Cuevas, and
  Pauly]{Kloeckner2016}
Kl{\"o}ckner,~J.~C.; B{\"u}rkle,~M.; Cuevas,~J.~C.; Pauly,~F. Length dependence
  of the thermal conductance of alkane-based single-molecule junctions: An
  \textit{ab initio} study. \emph{Phys. Rev. B} \textbf{2016}, \emph{94},
  205425--1--8\relax
\mciteBstWouldAddEndPuncttrue
\mciteSetBstMidEndSepPunct{\mcitedefaultmidpunct}
{\mcitedefaultendpunct}{\mcitedefaultseppunct}\relax
\EndOfBibitem
\bibitem[Wang \latin{et~al.}(2006)Wang, Segalman, and Majumdar]{Wang2006apl}
Wang,~R.~Y.; Segalman,~R.~A.; Majumdar,~A. Room temperature thermal conductance
  of alkanedithiol self-assembled monolayers. \emph{Appl. Phys. Lett.}
  \textbf{2006}, \emph{89}, 173113\relax
\mciteBstWouldAddEndPuncttrue
\mciteSetBstMidEndSepPunct{\mcitedefaultmidpunct}
{\mcitedefaultendpunct}{\mcitedefaultseppunct}\relax
\EndOfBibitem
\bibitem[Meier \latin{et~al.}(2014)Meier, Menges, Nirmalraj, H\"olscher, Riel,
  and Gotsmann]{Meier2014prl}
Meier,~T.; Menges,~F.; Nirmalraj,~P.; H\"olscher,~H.; Riel,~H.; Gotsmann,~B.
  Length-Dependent Thermal Transport along Molecular Chains. \emph{Phys. Rev.
  Lett.} \textbf{2014}, \emph{113}, 060801\relax
\mciteBstWouldAddEndPuncttrue
\mciteSetBstMidEndSepPunct{\mcitedefaultmidpunct}
{\mcitedefaultendpunct}{\mcitedefaultseppunct}\relax
\EndOfBibitem
\bibitem[Majumdar \latin{et~al.}(2015)Majumdar, Sierra-Suarez, Schiffres, Ong,
  Higgs, McGaughey, and Malen]{Majumdar2015nl}
Majumdar,~S.; Sierra-Suarez,~J.~A.; Schiffres,~S.~N.; Ong,~W.-L.; Higgs,~C.~F.;
  McGaughey,~A. J.~H.; Malen,~J.~A. Vibrational Mismatch of Metal Leads
  Controls Thermal Conductance of Self-Assembled Monolayer Junctions.
  \emph{Nano Letters} \textbf{2015}, \emph{15}, 2985--2991\relax
\mciteBstWouldAddEndPuncttrue
\mciteSetBstMidEndSepPunct{\mcitedefaultmidpunct}
{\mcitedefaultendpunct}{\mcitedefaultseppunct}\relax
\EndOfBibitem
\bibitem[Cui \latin{et~al.}(2019)Cui, Hur, Akbar, Kl{\"{o}}ckner, Jeong, Pauly,
  Jang, Reddy, and Meyhofer]{Cui2019nature}
Cui,~L.; Hur,~S.; Akbar,~Z.~A.; Kl{\"{o}}ckner,~J.~C.; Jeong,~W.; Pauly,~F.;
  Jang,~S.-Y.; Reddy,~P.; Meyhofer,~E. {Thermal conductance of single-molecule
  junctions}. \emph{Nature} \textbf{2019}, \relax
\mciteBstWouldAddEndPunctfalse
\mciteSetBstMidEndSepPunct{\mcitedefaultmidpunct}
{}{\mcitedefaultseppunct}\relax
\EndOfBibitem
\bibitem[Crljen and Baranovi{\'{c}}(2007)Crljen, and
  Baranovi{\'{c}}]{Crljen2007prl}
Crljen,~{\v{Z}}.; Baranovi{\'{c}},~G. {Unusual conductance of polyyne-based
  molecular wires}. \emph{Phys. Rev. Lett.} \textbf{2007}, \emph{98},
  1--4\relax
\mciteBstWouldAddEndPuncttrue
\mciteSetBstMidEndSepPunct{\mcitedefaultmidpunct}
{\mcitedefaultendpunct}{\mcitedefaultseppunct}\relax
\EndOfBibitem
\bibitem[Garner \latin{et~al.}(2018)Garner, Bro-J{\o}rgensen, Pedersen, and
  Solomon]{Garner2018jpcc}
Garner,~M.~H.; Bro-J{\o}rgensen,~W.; Pedersen,~P.~D.; Solomon,~G.~C. {Reverse
  Bond-Length Alternation in Cumulenes: Candidates for Increasing Electronic
  Transmission with Length}. \emph{J. Phys. Chem. C} \textbf{2018},
  acs.jpcc.8b05661\relax
\mciteBstWouldAddEndPuncttrue
\mciteSetBstMidEndSepPunct{\mcitedefaultmidpunct}
{\mcitedefaultendpunct}{\mcitedefaultseppunct}\relax
\EndOfBibitem
\bibitem[Torii \latin{et~al.}(2008)Torii, Nakano, and Ohara]{Torii2008jcp}
Torii,~D.; Nakano,~T.; Ohara,~T. {Contribution of inter- and intramolecular
  energy transfers to heat conduction in liquids}. \emph{J. Chem. Phys.}
  \textbf{2008}, \emph{128}\relax
\mciteBstWouldAddEndPuncttrue
\mciteSetBstMidEndSepPunct{\mcitedefaultmidpunct}
{\mcitedefaultendpunct}{\mcitedefaultseppunct}\relax
\EndOfBibitem
\bibitem[Pronk \latin{et~al.}(2013)Pronk, Páll, Schulz, Larsson, Bjelkmar,
  Apostolov, Shirts, Smith, Kasson, van~der Spoel, Hess, and
  Lindahl]{Gromacs4.5}
Pronk,~S.; Páll,~S.; Schulz,~R.; Larsson,~P.; Bjelkmar,~P.; Apostolov,~R.;
  Shirts,~M.~R.; Smith,~J.~C.; Kasson,~P.~M.; van~der Spoel,~D.; Hess,~B.;
  Lindahl,~E. GROMACS 4.5: a high-throughput and highly parallel open source
  molecular simulation toolkit. \emph{Bioinformatics} \textbf{2013}, \emph{29},
  845--854\relax
\mciteBstWouldAddEndPuncttrue
\mciteSetBstMidEndSepPunct{\mcitedefaultmidpunct}
{\mcitedefaultendpunct}{\mcitedefaultseppunct}\relax
\EndOfBibitem
\bibitem[O'Boyle \latin{et~al.}(2011)O'Boyle, Banck, James, Morley,
  Vandermeersch, and Hutchison]{OpenBabel}
O'Boyle,~N.~M.; Banck,~M.; James,~C.~A.; Morley,~C.; Vandermeersch,~T.;
  Hutchison,~G.~R. Open Babel: An open chemical toolbox. \emph{J.
  Cheminformatics} \textbf{2011}, \emph{3}, 33\relax
\mciteBstWouldAddEndPuncttrue
\mciteSetBstMidEndSepPunct{\mcitedefaultmidpunct}
{\mcitedefaultendpunct}{\mcitedefaultseppunct}\relax
\EndOfBibitem
\bibitem[Hanwell \latin{et~al.}(2012)Hanwell, Curtis, Lonie, Vandermeersch,
  Zurek, and Hutchison]{Avogadro}
Hanwell,~M.~D.; Curtis,~D.~E.; Lonie,~D.~C.; Vandermeersch,~T.; Zurek,~E.;
  Hutchison,~G.~R. Avogadro: an advanced semantic chemical editor,
  visualization, and analysis platform. \emph{Journal of Cheminformatics}
  \textbf{2012}, \emph{4}, 17\relax
\mciteBstWouldAddEndPuncttrue
\mciteSetBstMidEndSepPunct{\mcitedefaultmidpunct}
{\mcitedefaultendpunct}{\mcitedefaultseppunct}\relax
\EndOfBibitem
\bibitem[Rappe \latin{et~al.}(1992)Rappe, Casewit, Colwell, Goddard, and
  Skiff]{UFF}
Rappe,~A.~K.; Casewit,~C.~J.; Colwell,~K.~S.; Goddard,~W.~A.; Skiff,~W.~M. UFF,
  a full periodic table force field for molecular mechanics and molecular
  dynamics simulations. \emph{Journal of the American Chemical Society}
  \textbf{1992}, \emph{114}, 10024--10035\relax
\mciteBstWouldAddEndPuncttrue
\mciteSetBstMidEndSepPunct{\mcitedefaultmidpunct}
{\mcitedefaultendpunct}{\mcitedefaultseppunct}\relax
\EndOfBibitem
\bibitem[Zhang \latin{et~al.}(2010)Zhang, Barnes, Yan, and Hase]{Zhang2010pccp}
Zhang,~Y.; Barnes,~G.~L.; Yan,~T.; Hase,~W.~L. Model non-equilibrium molecular
  dynamics simulations of heat transfer from a hot gold surface to an
  alkylthiolate self-assembled monolayer. \emph{Phys. Chem. Chem. Phys.}
  \textbf{2010}, \emph{12}, 4435--4445\relax
\mciteBstWouldAddEndPuncttrue
\mciteSetBstMidEndSepPunct{\mcitedefaultmidpunct}
{\mcitedefaultendpunct}{\mcitedefaultseppunct}\relax
\EndOfBibitem
\bibitem[Carles \latin{et~al.}(2016)Carles, Benzo, P{\'{e}}cassou, and
  Bonafos]{Carles2016scirep}
Carles,~R.; Benzo,~P.; P{\'{e}}cassou,~B.; Bonafos,~C. {Vibrational density of
  states and thermodynamics at the nanoscale: the 3D-2D transition in gold
  nanostructures}. \emph{Sci. Rep.} \textbf{2016}, \emph{6}, 39164\relax
\mciteBstWouldAddEndPuncttrue
\mciteSetBstMidEndSepPunct{\mcitedefaultmidpunct}
{\mcitedefaultendpunct}{\mcitedefaultseppunct}\relax
\EndOfBibitem
\bibitem[Altman and Bland(2005)Altman, and Bland]{SE}
Altman,~D.~G.; Bland,~J.~M. {Standard deviations and standard errors}.
  \emph{BMJ (Clinical research ed.)} \textbf{2005}, \emph{331}, 903\relax
\mciteBstWouldAddEndPuncttrue
\mciteSetBstMidEndSepPunct{\mcitedefaultmidpunct}
{\mcitedefaultendpunct}{\mcitedefaultseppunct}\relax
\EndOfBibitem
\end{mcitethebibliography}


\providecommand{\latin}[1]{#1}
\makeatletter
\providecommand{\doi}
  {\begingroup\let\do\@makeother\dospecials
  \catcode`\{=1 \catcode`\}=2 \doi@aux}
\providecommand{\doi@aux}[1]{\endgroup\texttt{#1}}
\makeatother
\providecommand*\mcitethebibliography{\thebibliography}
\csname @ifundefined\endcsname{endmcitethebibliography}
  {\let\endmcitethebibliography\endthebibliography}{}
\begin{mcitethebibliography}{13}
\providecommand*\natexlab[1]{#1}
\providecommand*\mciteSetBstSublistMode[1]{}
\providecommand*\mciteSetBstMaxWidthForm[2]{}
\providecommand*\mciteBstWouldAddEndPuncttrue
  {\def\EndOfBibitem{\unskip.}}
\providecommand*\mciteBstWouldAddEndPunctfalse
  {\let\EndOfBibitem\relax}
\providecommand*\mciteSetBstMidEndSepPunct[3]{}
\providecommand*\mciteSetBstSublistLabelBeginEnd[3]{}
\providecommand*\EndOfBibitem{}
\mciteSetBstSublistMode{f}
\mciteSetBstMaxWidthForm{subitem}{(\alph{mcitesubitemcount})}
\mciteSetBstSublistLabelBeginEnd
  {\mcitemaxwidthsubitemform\space}
  {\relax}
  {\relax}

\bibitem[Pronk \latin{et~al.}(2013)Pronk, Páll, Schulz, Larsson, Bjelkmar,
  Apostolov, Shirts, Smith, Kasson, van~der Spoel, Hess, and
  Lindahl]{Gromacs4.5}
Pronk,~S.; Páll,~S.; Schulz,~R.; Larsson,~P.; Bjelkmar,~P.; Apostolov,~R.;
  Shirts,~M.~R.; Smith,~J.~C.; Kasson,~P.~M.; van~der Spoel,~D.; Hess,~B.;
  Lindahl,~E. GROMACS 4.5: a high-throughput and highly parallel open source
  molecular simulation toolkit. \emph{Bioinformatics} \textbf{2013}, \emph{29},
  845--854\relax
\mciteBstWouldAddEndPuncttrue
\mciteSetBstMidEndSepPunct{\mcitedefaultmidpunct}
{\mcitedefaultendpunct}{\mcitedefaultseppunct}\relax
\EndOfBibitem
\bibitem[Hanwell \latin{et~al.}(2012)Hanwell, Curtis, Lonie, Vandermeersch,
  Zurek, and Hutchison]{Avogadro}
Hanwell,~M.~D.; Curtis,~D.~E.; Lonie,~D.~C.; Vandermeersch,~T.; Zurek,~E.;
  Hutchison,~G.~R. Avogadro: an advanced semantic chemical editor,
  visualization, and analysis platform. \emph{Journal of Cheminformatics}
  \textbf{2012}, \emph{4}, 17\relax
\mciteBstWouldAddEndPuncttrue
\mciteSetBstMidEndSepPunct{\mcitedefaultmidpunct}
{\mcitedefaultendpunct}{\mcitedefaultseppunct}\relax
\EndOfBibitem
\bibitem[O'Boyle \latin{et~al.}(2011)O'Boyle, Banck, James, Morley,
  Vandermeersch, and Hutchison]{OpenBabel}
O'Boyle,~N.~M.; Banck,~M.; James,~C.~A.; Morley,~C.; Vandermeersch,~T.;
  Hutchison,~G.~R. Open Babel: An open chemical toolbox. \emph{J.
  Cheminformatics} \textbf{2011}, \emph{3}, 33\relax
\mciteBstWouldAddEndPuncttrue
\mciteSetBstMidEndSepPunct{\mcitedefaultmidpunct}
{\mcitedefaultendpunct}{\mcitedefaultseppunct}\relax
\EndOfBibitem
\bibitem[Rappe \latin{et~al.}(1992)Rappe, Casewit, Colwell, Goddard, and
  Skiff]{UFF}
Rappe,~A.~K.; Casewit,~C.~J.; Colwell,~K.~S.; Goddard,~W.~A.; Skiff,~W.~M. UFF,
  a full periodic table force field for molecular mechanics and molecular
  dynamics simulations. \emph{Journal of the American Chemical Society}
  \textbf{1992}, \emph{114}, 10024--10035\relax
\mciteBstWouldAddEndPuncttrue
\mciteSetBstMidEndSepPunct{\mcitedefaultmidpunct}
{\mcitedefaultendpunct}{\mcitedefaultseppunct}\relax
\EndOfBibitem
\bibitem[Segal and Agarwalla(2016)Segal, and Agarwalla]{Segal2016arpc}
Segal,~D.; Agarwalla,~B.~K. Vibrational Heat Transport in Molecular Junctions.
  \emph{Annu. Rev. Phys. Chem.} \textbf{2016}, \emph{67}, 185--209\relax
\mciteBstWouldAddEndPuncttrue
\mciteSetBstMidEndSepPunct{\mcitedefaultmidpunct}
{\mcitedefaultendpunct}{\mcitedefaultseppunct}\relax
\EndOfBibitem
\bibitem[Dhar and Roy(2006)Dhar, and Roy]{Dhar2006}
Dhar,~A.; Roy,~D. Heat Transport in Harmonic Lattices. \emph{J. Stat. Phys.}
  \textbf{2006}, \emph{125}, 801--820\relax
\mciteBstWouldAddEndPuncttrue
\mciteSetBstMidEndSepPunct{\mcitedefaultmidpunct}
{\mcitedefaultendpunct}{\mcitedefaultseppunct}\relax
\EndOfBibitem
\bibitem[Dhar(2008)]{Dhar2008}
Dhar,~A. Heat transport in low-dimensional systems. \emph{Adv. Phys.}
  \textbf{2008}, \emph{57}, 457--537\relax
\mciteBstWouldAddEndPuncttrue
\mciteSetBstMidEndSepPunct{\mcitedefaultmidpunct}
{\mcitedefaultendpunct}{\mcitedefaultseppunct}\relax
\EndOfBibitem
\bibitem[Segal \latin{et~al.}(2003)Segal, Nitzan, and H\"anggi]{Segal2003}
Segal,~D.; Nitzan,~A.; H\"anggi,~P. Thermal conductance through molecular
  wires. \emph{J. Chem. Phys.} \textbf{2003}, \emph{119}, 6840--6855\relax
\mciteBstWouldAddEndPuncttrue
\mciteSetBstMidEndSepPunct{\mcitedefaultmidpunct}
{\mcitedefaultendpunct}{\mcitedefaultseppunct}\relax
\EndOfBibitem
\bibitem[Wang \latin{et~al.}(2006)Wang, Wang, and Zeng]{Wang2006prb}
Wang,~J.-S.; Wang,~J.; Zeng,~N. Nonequilibrium Green's function approach to
  mesoscopic thermal transport. \emph{Phys. Rev. B} \textbf{2006}, \emph{74},
  033408\relax
\mciteBstWouldAddEndPuncttrue
\mciteSetBstMidEndSepPunct{\mcitedefaultmidpunct}
{\mcitedefaultendpunct}{\mcitedefaultseppunct}\relax
\EndOfBibitem
\bibitem[Yamamoto and Watanabe(2006)Yamamoto, and Watanabe]{Yamamoto2006}
Yamamoto,~T.; Watanabe,~K. Nonequilibrium Green's Function Approach to Phonon
  Transport in Defective Carbon Nanotubes. \emph{Phys. Rev. Lett.}
  \textbf{2006}, \emph{96}, 255503\relax
\mciteBstWouldAddEndPuncttrue
\mciteSetBstMidEndSepPunct{\mcitedefaultmidpunct}
{\mcitedefaultendpunct}{\mcitedefaultseppunct}\relax
\EndOfBibitem
\bibitem[Kl{\"o}ckner \latin{et~al.}(2016)Kl{\"o}ckner, B{\"u}rkle, Cuevas, and
  Pauly]{Kloeckner2016}
Kl{\"o}ckner,~J.~C.; B{\"u}rkle,~M.; Cuevas,~J.~C.; Pauly,~F. Length dependence
  of the thermal conductance of alkane-based single-molecule junctions: An
  \textit{ab initio} study. \emph{Phys. Rev. B} \textbf{2016}, \emph{94},
  205425--1--8\relax
\mciteBstWouldAddEndPuncttrue
\mciteSetBstMidEndSepPunct{\mcitedefaultmidpunct}
{\mcitedefaultendpunct}{\mcitedefaultseppunct}\relax
\EndOfBibitem
\bibitem[Meir and Wingreen(1992)Meir, and Wingreen]{Meir1992prl}
Meir,~Y.; Wingreen,~N.~S. Landauer formula for the current through an
  interacting electron region. \emph{Phys. Rev. Lett.} \textbf{1992},
  \emph{68}, 2512--2515\relax
\mciteBstWouldAddEndPuncttrue
\mciteSetBstMidEndSepPunct{\mcitedefaultmidpunct}
{\mcitedefaultendpunct}{\mcitedefaultseppunct}\relax
\EndOfBibitem
\end{mcitethebibliography}

\end{document}


\section{S1. Details on molecular configuration preparation and MD simulation procedures}
There are mainly three phases in using our customized GROMACS\cite{Gromacs4.5} package
to conduct simulations on heat conduction in molecular junctions, 
including both classical MD and quantum Landauer transport 
(as illustrated in figure \ref{fig:procedure}):

\begin{figure}[H]
    \centering
\includegraphics[width=1\textwidth]{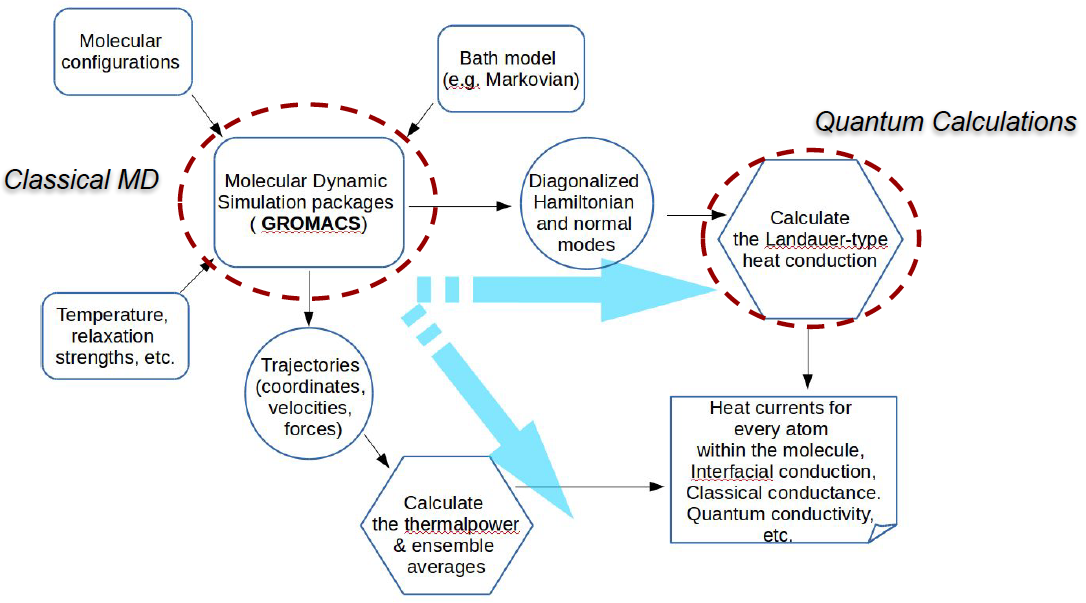}
    \caption{Molecular heat conduction MD simulation procedure and implementation diagram,
    		with Landauer-type quantum calculations included.}
    \label{fig:procedure}
\end{figure}

(I) Configuration preparation. 
The open-source graphic cross-platform molecule editor, 
\textit{Avogardro}\cite{Avogadro},
is used to design the initial molecular topologies as one wishes
(e.g. alkanedithiols with three layers of gold substrates).
Then these configurations are energy minimized 
with \textit{Avogardro}'s native utility,
and saved as \textit{pdb} formats,
which are eventually transformed into GROMACS's recognizable input files
(\textit{gro} and \textit{top})
using \textit{Open Babel}\cite{OpenBabel}
with Uinversal Force Field (UFF)\cite{UFF} 
implemented as molecular force field parameters.

(II) Production runs.
Within GROMACS the molecualr configurations are further energy minimized
to ensure stable structure for subsequent dynamical runs and normal mode analysis.
For MD production runs, the systems are first equilibrated to the average temperatures
of the baths (usually takes a few hundred pecoseconds), 
and then with a single long trajectory (a few nanoseconds) relaxed 
to nonequilibrium steady state. 
From the tail of this trajectory, thousands of parallel
at-steady-state trajectories are launched
get large enough statisical ensembles.
At the end of this phase, as outputs, normal modes, steady-state trajectories,
and inter-atomic forces are collected.

(III) Post-data processing.
The heat currents are calculated according to the equation
in the Main text using velocities and inter-atomic
forces sampled at the end of the last step.
The Landauer-type currents are calcualted 
through a different channel, 
by using the formalism described in the last section of this SI.
The resultant quantities (e.g. currents, local temperatures, conductance, \textit{etc.})
are averaged (with error estimation) and plotted for illustration purposes.

\section{S2. Supplemental MD simulation results}

\begin{figure}[H]
    \centering
\includegraphics[width=1\textwidth]{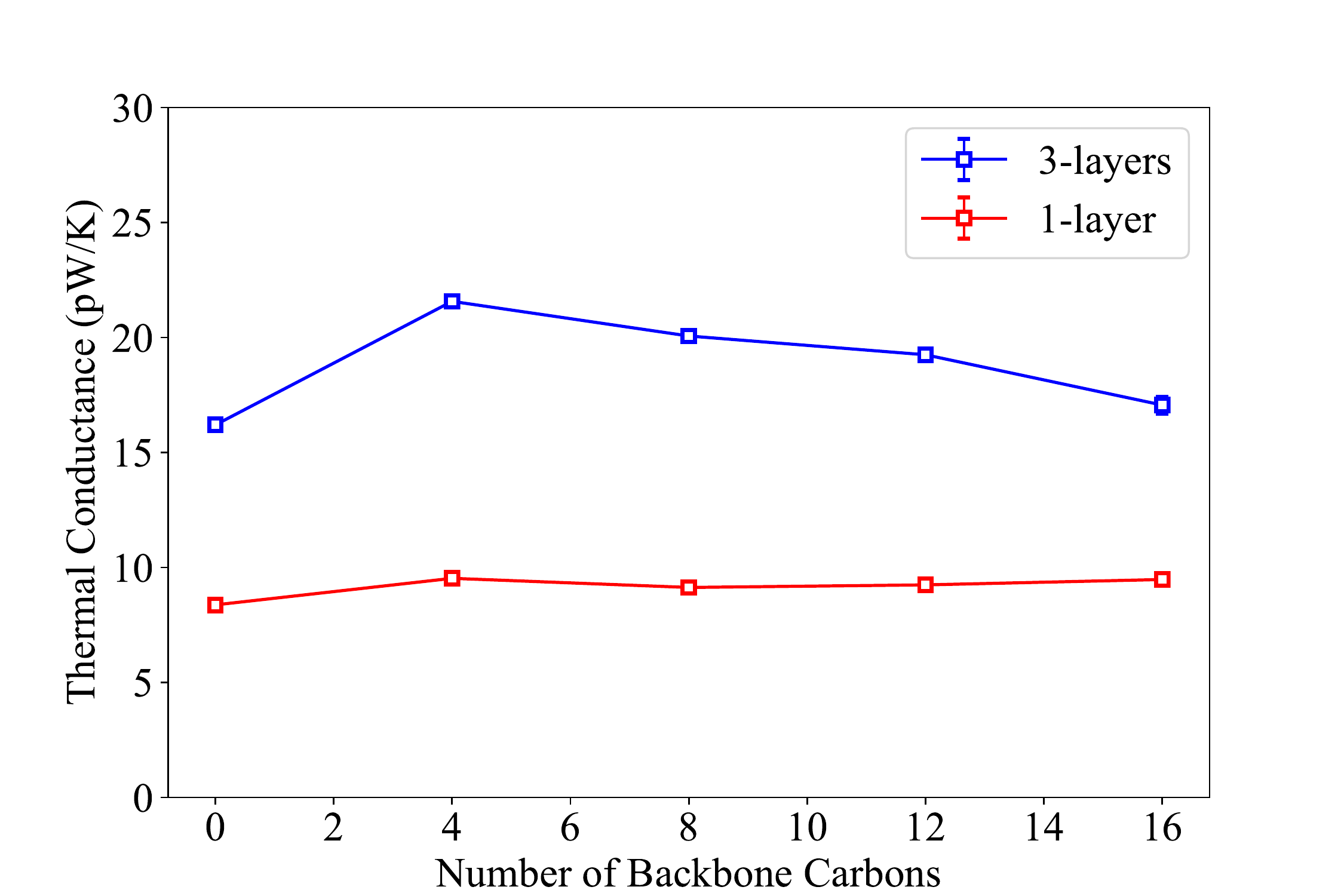}
    \caption{Heat conductance simulated for alkanes of various lengths, under explicit baths of different number of explicit Gold layers.
 The bath temperatures are set at 300K to 350K. The bars shown in the figure are the standard errors.}
    \label{fig:AuLayer}
\end{figure}

\begin{figure}[H]
    \centering
    \includegraphics[width=1\textwidth]{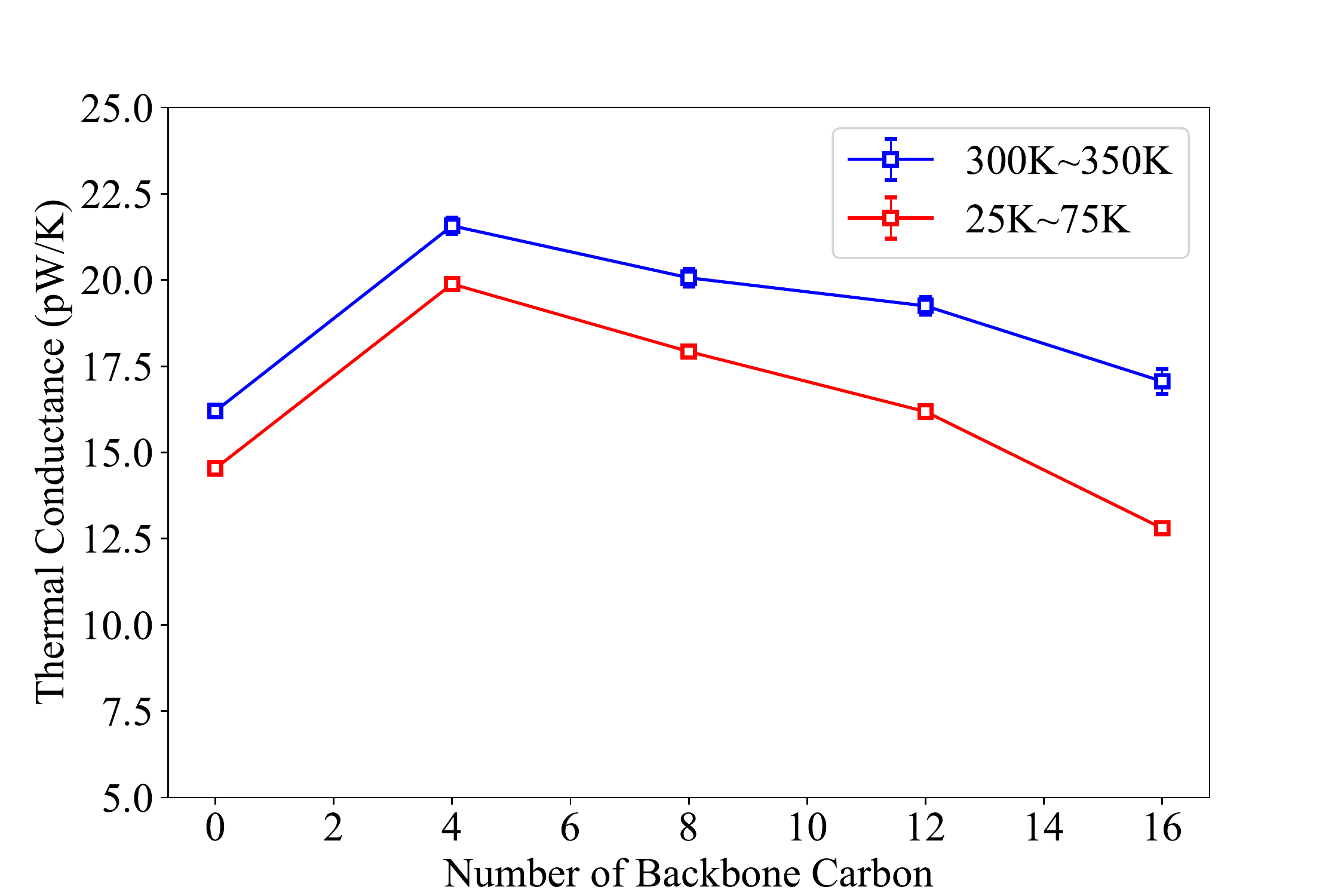}
    \caption{Comparison of low temperature conductance with room temperature conductance for different length of alkanedithiols with 3-layers of explicit gold bulks, from GROMACS MD simulations.}
    \label{fig:conductance_alkanes_lowTvsRoomT_upto16}
\end{figure}

\section{S3. Landauer-type calculations for hydrocarbon heat conduction}
\subsection{Landauer formalism}
When only the harmonic part of the system-bath interactions are taken into account,
the (phononic) heat current can be expressed by the Landauer formula, 
\begin{align}
    \label{eqn:Landauer_heat}
    J = \frac { \hbar } { 2\pi } \int _ {0} ^ { \infty } \mathcal { T } ( \omega )  \left[ f \left( \omega , T _ { L } \right) - f \left( \omega , T _ { R } \right) \right] \omega d \omega 
\end{align}
which has been derived in various theoretical approaches and employed in different systems\cite{Segal2016arpc,Dhar2006,Dhar2008,Segal2003,Wang2006prb,Yamamoto2006,Kloeckner2016}. Here $f$ function is the Bose-Einstein distribution function which depends the temperature of the bath, $f(\omega, T)=(e^{\hbar\omega/k_BT}-1)^{-1}$, 
and $\mathcal { T }$ is the transmission probability.

In the Linear Response regime,
the thermal conductance is taken as the derivative of the heat current with respect to the temperature,
\begin{align}
    \kappa=\frac { \hbar } { 2\pi } \int _ {0} ^ { \infty }  \mathcal { T } ( \omega ) \frac{\partial f \left( \omega , T  \right)}{\partial T} \omega d \omega
\end{align}

The transmission function is often expressed using the Meir-Wingreen formula\cite{Meir1992prl}
\begin{align}
    \label{eqn:transmission}
    \mathcal { T } ( \omega ) = \operatorname { Tr } \left[ \boldsymbol { G } _ { S } ^ { r } ( \omega ) \mathbf { \Gamma } _ { L } ( \omega ) \boldsymbol { G } _ { S } ^ { a } ( \omega ) \mathbf { \Gamma } _ { R } ( \omega ) \right],
\end{align}
in which $\bm{G}_S^a=[\bm{G}_S^r]^\dagger$, are the advanced \& retarded Green's functions of the system and can be written as\cite{Dhar2006},
\begin{align}
	\label{eqn:Green_harmonic}
	\bm{G}_S^{r/a}(\omega)=[\omega^2\bm{M}-\bm{D}-(\bm{\Sigma}_L^{r/a}+\bm{\Sigma}_R^{r/a})]^{-1},
\end{align}

most easily understood in the basis of the atomic coordinates,  $\bm{D}$ is the dynamical matrix (or Hessian) whose elements are the second derivatives of the total energy with respect to the atomic coordinates, 
$\bm{M}$ is the (diagonal) matrix of atomic masses, and $\mathbf{\Sigma}^{r/a}_K$ is the retarded (or advanced, respectively) self-energy of the $K$ bath ($K \in {L, R}$).

$\mathbf{\Gamma}_L$ and $\mathbf{\Gamma}_R$ are the spectral function matrices of the baths, reflecting the coupling strengths of the systems to the leads.
\begin{align}
	\label{eqn:coupling}
	\bm{\Gamma}_{L/R}(\omega)=i[\bm{\Sigma}^r_{L/R}(\omega)-\bm{\Sigma}^a_{L/R}(\omega)].
\end{align}

When spectral density is taken as Ohmic\cite{Dhar2008,Segal2016arpc}, the self-energies are defined as $[\bm{\Sigma}^r_L]_{ij}\equiv -i\omega\gamma_L(\omega)\delta_{ij}$ for all atomic indexes $i$ (and $j$) which correspond to coordinates of atoms at the left-most layer of bulk, and directly coupled to the left heat reservoir, which is characterised by a dissipation coefficient $\gamma_L(\omega)$. Correspondingly for the right bath.
Also, taking both baths to be white, $\gamma_L(\omega)$ and $\gamma_R(\omega)$ are constant.

\subsection{Methodology}
The main ingredient needed for the use of the Landauer formula in molecular systems 
is to reliably obtain the dynamical force matrix $\bm{D}$.
The GROMACS~\cite{Gromacs4.5} software package provides utilities which construct the Hessian matrix of the system under study. Furthermore, it has the ability to diagonalize the obtained Hessian, providing the eigenfrequencies and eigenvectors (normal modes) of the system.

\subsection{Results}

\begin{figure}[H]
    \centering
    \includegraphics[width=1\textwidth]{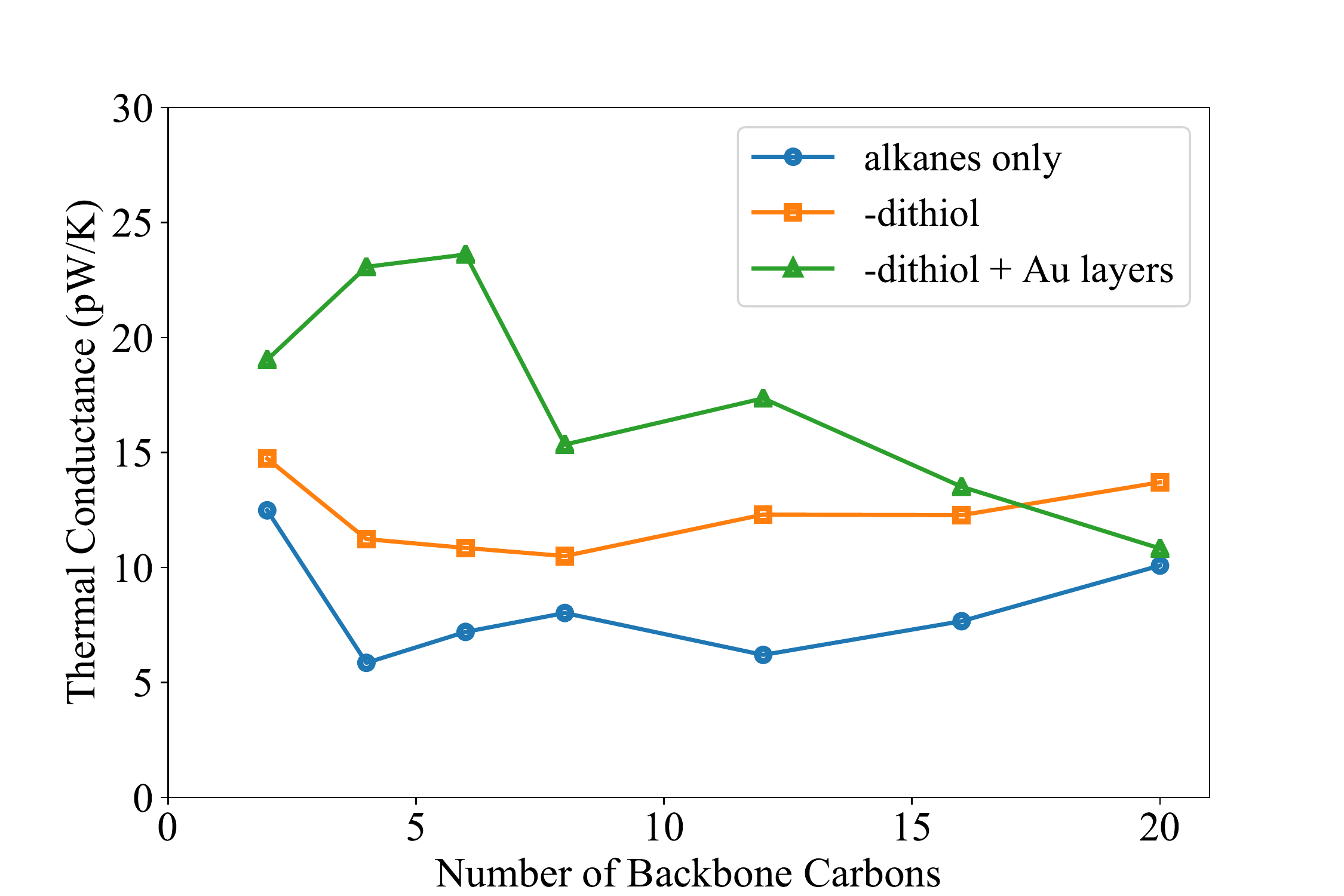}
    \caption{Length-dependent thermal conductance, calculated through the Landauer formula. The exterior-most layers of explicit bulk are coupled to white noise baths. The conductance denotes ratio between heat current and temperature bias, with left and right baths at 300K and 350K respectively. There are three Gold layers,
    in accordance with the MD simulations.}
    \label{fig:conductance_Landauer_20}
\end{figure}

By adding layers to different alkane molecules(Figure \ref{fig:conductance_Landauer_20}), we find

\begin{enumerate}
\item The presence of thiol groups enhances alkane conductance.
\item The trend of conductance increase and eventual decrease (as a function of chain length), as reported in \textit{ab-initio} calculations of similar systems\cite{Kloeckner2016}, appears only when including the layers of Gold leads, explicitly.
\end{enumerate}

To see how different molecular vibrational modes line up with the bath spectral density, the Normal Mode Analysis (NMA) of various molecules is performed (results shown in Figure \ref{fig:modes_histogram_triple}).

\begin{figure}[H]
    \centering
    \includegraphics[width=1\textwidth]{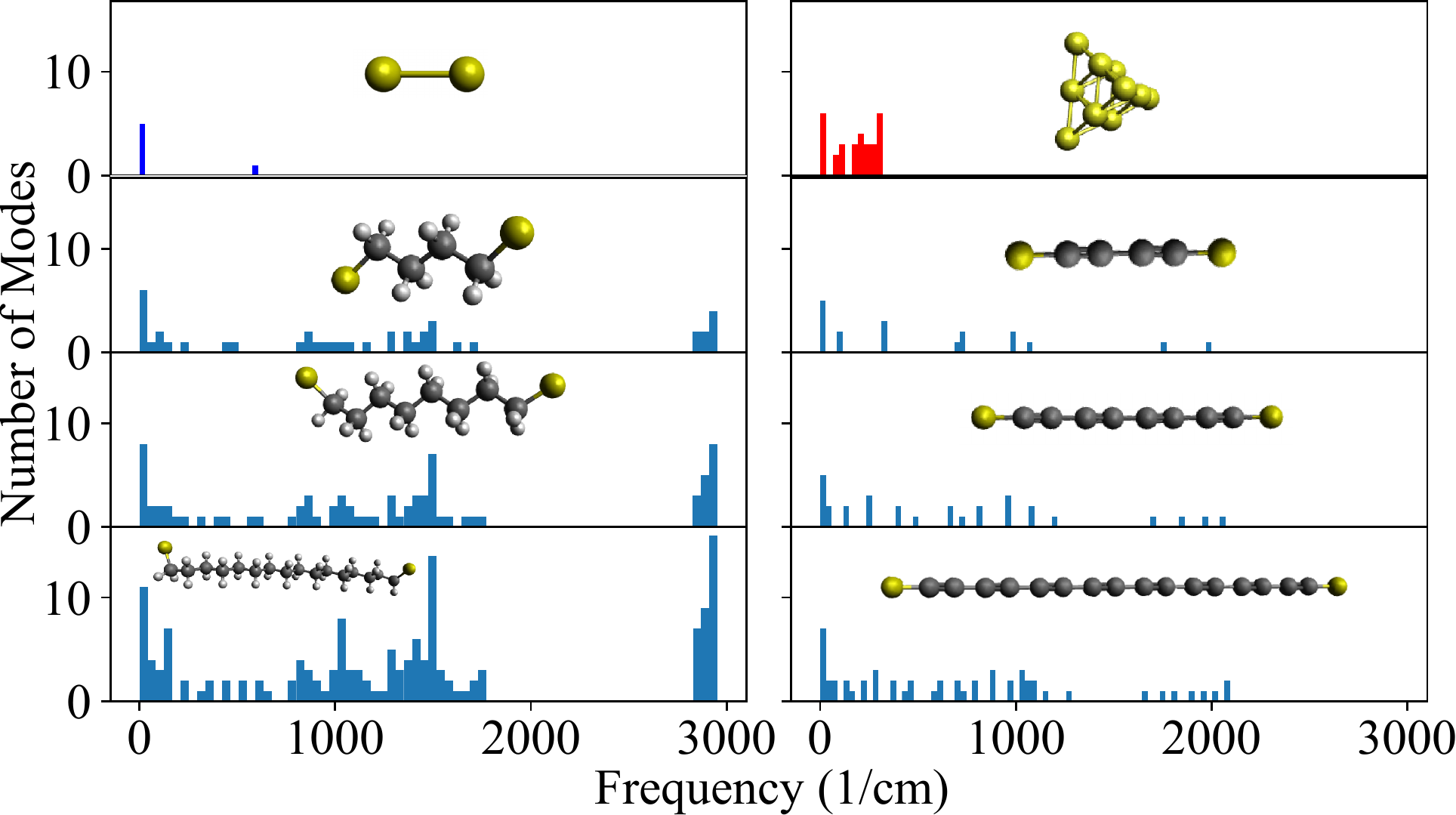}
    \caption{
    Histograms of the Normal Mode Analysis of some representative molecules.
    The x-axes are the histogram bins (frequencies in wavenumbers),and the y-axes are the count of modes with frequency in the bin.
    The corresponding molecular topologies are drawn within each panel.
    Note that the first and last atom in each molecule are Sulfur.
    The first row shows the normal mode histogram for disulfur and a three layer gold cluster (left and right columns, respectively).
    The following rows depict alkanes (left) and polyyens (right) with 4, 8, and 16 Carbon units, respectively.}
    \label{fig:modes_histogram_triple}
\end{figure}

While the molecules have some modes up to a thousand wavenumbers, the significant band of bulk phonons exists between a few tens and a few hundred wavenumbers, 
as seen in gold bulk spetrum density figure in the main text.
The high-frequency molecular vibrational modes, therefore, do not overlap with the bulk modes.
The prominent peaks of molecular vibrations at lower frequencies are due to rigid-body Center of Mass Motion, or such Degrees of Freedom as they are hybridized with vibrations.
Other factors which could be affecting the overall heat conduction include, normal mode localization\cite{Segal2003}, system-bath coupling, transmission coefficients in the picture of coherent transport\cite{Kloeckner2016}.
The relevant investigations will require not only analysis of energy flows within the molecules and between molecules and baths, but also in frequency space, and detailed comparisons to Landauer-type conductivity calculations. 

We compare the localization properties of the normal modes 
for the molecules of different lengths and saturations.
Assuming the transformation coefficients for 
changing the local atomic basis to normal mode basis to be $C_{n_i,\alpha}$,
then a localization factor (also called participation index) for each mode can be defined 
as \cite{Segal2003}:

\begin{align}
	\label{}
	P_\alpha(n) = \sum_{n_i} |C_{n_i,\alpha}|^2,
\end{align}
where n is the index of atoms in the molecule, $\alpha$ denote the modes,
$n_i$ is the degrees of freedom which goes from 1 to 3 for each n.
To see the localization on carbon based backbone clusters, 
the factor can further be summed up over nearby hydrogens.
\begin{align}
	P_\alpha(n_B) = \sum_{n=1}^{n_H}P_\alpha(n),
\end{align}
where $n_H$ denotes the bonded hydrogens to backbone atom $n_B$.
\begin{figure}\ContinuedFloat
    \includegraphics[width=1\textwidth]{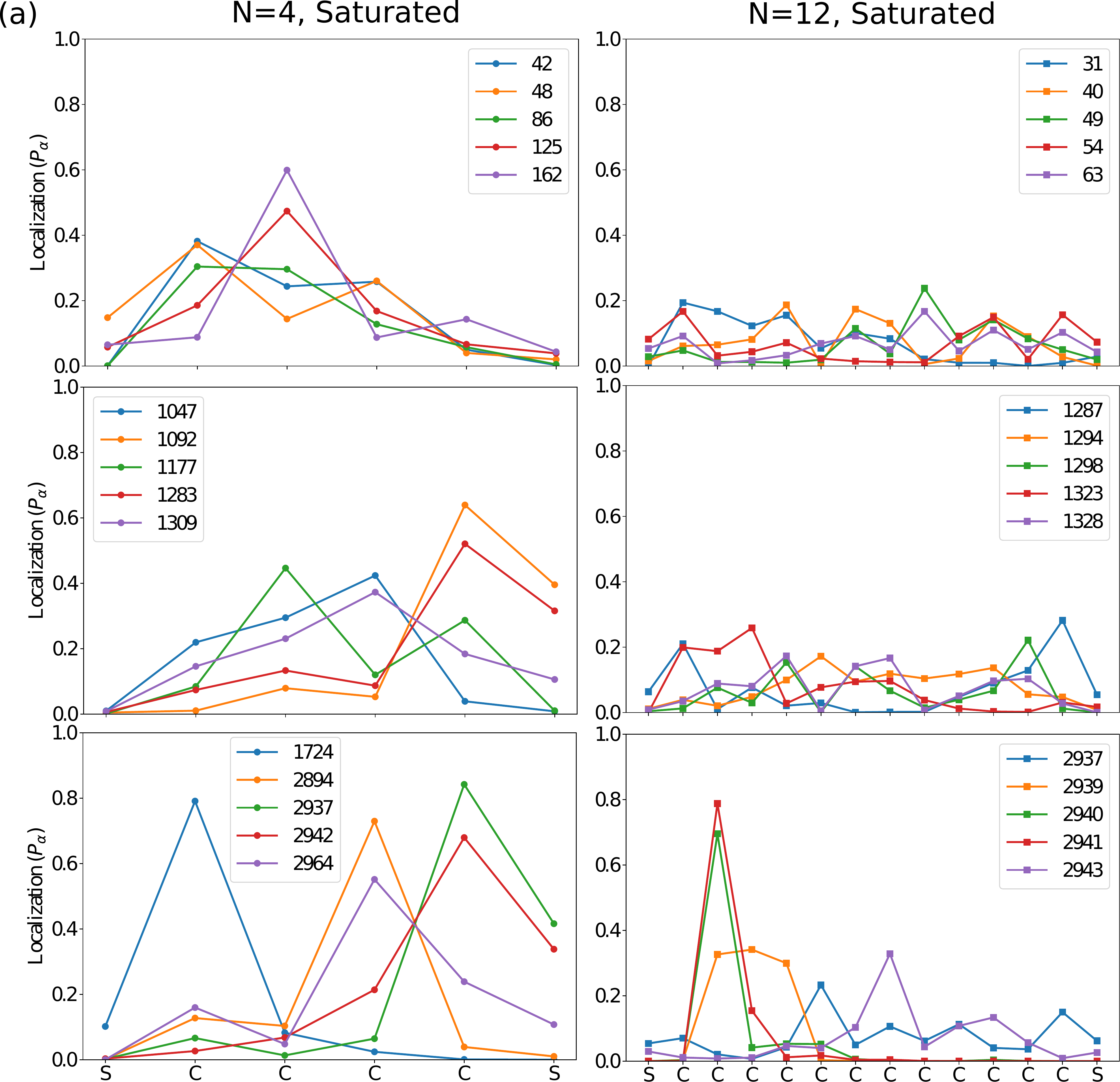}
\end{figure}
\begin{figure}[H]
    \centering
    \includegraphics[width=1\textwidth]{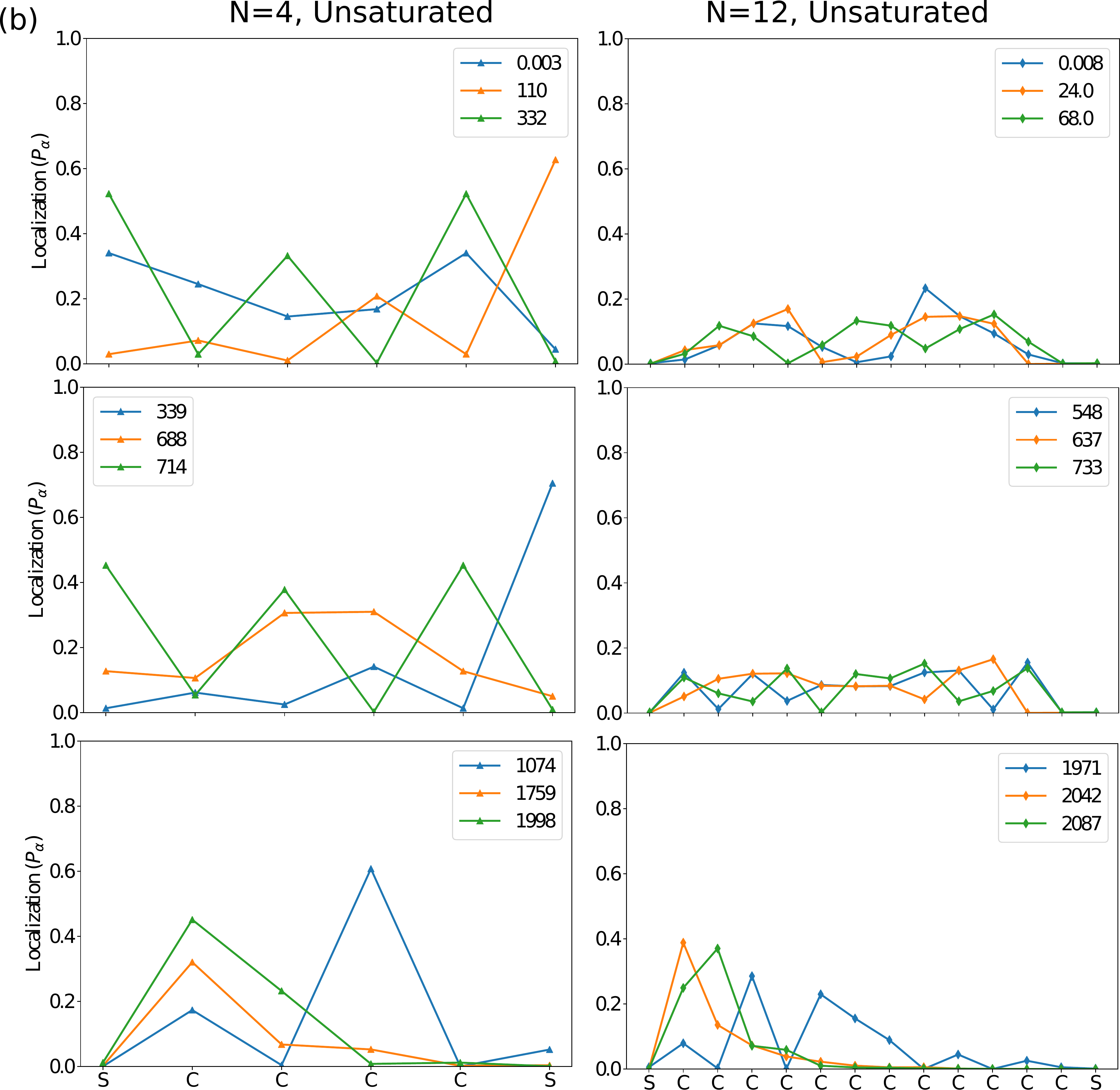}
    \caption{Localization coefficients
    for different hydrocarbon molecules. 
    $N$ is the number of backbone carbon atoms, and all molecules are capped with thiol groups. 
    "Saturated" refers to alkanedithiols(upper six panels labeled as (a)), 
    while "Unsaturated" refers to conjugated carbon chains with alternating single and triple bond (Polyynes)(lower six pannels labeled as (b)). 
    The numbers in the legends are the corresponding frequencies (unit in $cm^{-1}$, ranging from low, medium to high) 
    of the selected normal modes of the hydrocarbon chain molecules. 
    }
    \label{fig:localization}
\end{figure}
From figure \ref{fig:localization}, the high frequency modes
($\sim 3000 cm^{-1}$ for alkanes and $\sim 2000 cm^{-1}$ for polyynes) seem 
more spatially localized than low (less than 100 $cm^{-1}$) and medium
($\sim 1200 cm^{-1}$ for alkanes and $\sim 600 cm^{-1}$ for polyynes)
frequency modes, which might indicate that those high frequency modes
are less conductive even when the baths contain high frequency phonons.
Nonetheless, one has to be careful not to ignore
the couplings among molecular modes and between system and baths as well.

\begin{figure}[H]
    \centering
    \includegraphics[width=1\textwidth]{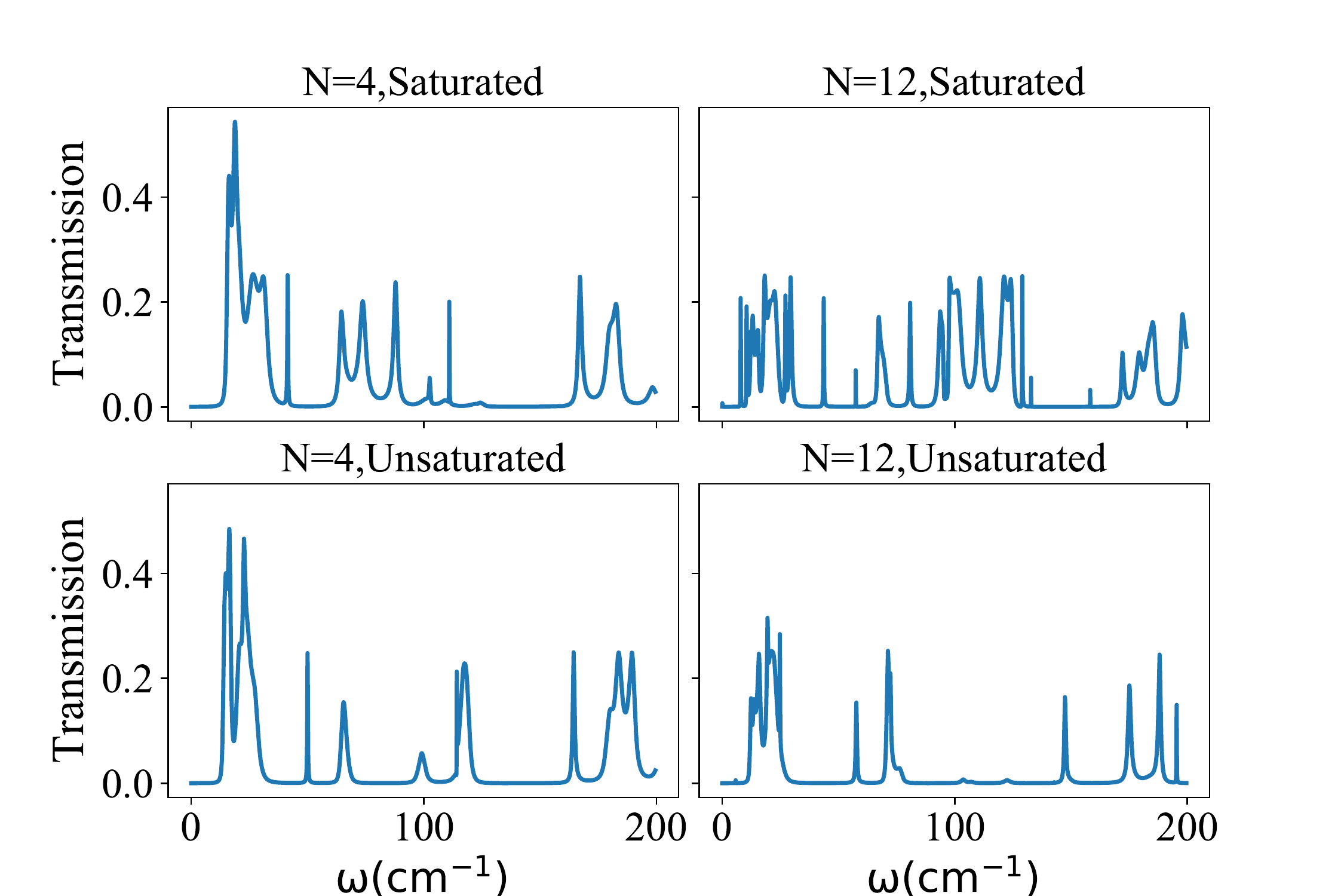}
    \caption{The transmission probabilities 
    $ \mathcal{T}(\omega) $
    for different hydrocarbon molecules. 
    $N$ is the number of backbone carbon atoms, and all molecules are capped with thiol groups. 
    "Saturated" refers to alkanedithiols, while "Unsaturated" refers to conjugated carbon chains with alternating single and triple bond (Polyynes). 
    Each molecule is adsorbed on a gold surface modeled using 3 explicit bulk layers.}
    \label{fig:transmission}
\end{figure}

As an important component in Landuaer's description,
transmission coefficients may become more a direct indicator 
of heat conduction compared to normal mode density and localization.
The comparison of transmission between short and long chains, of both saturated and unsaturated hydrocarbons (Figure \ref{fig:transmission}), provides insights into the contribution of different vibrational modes
to the overall heat conduction. Based on these results, one may make a qualitative observation that the short saturated hydrocarbons are the most conducting, while long unsaturated hydrocarbons are less conductive.

\begin{figure}[H]
    \centering
    \includegraphics[width=1\textwidth]{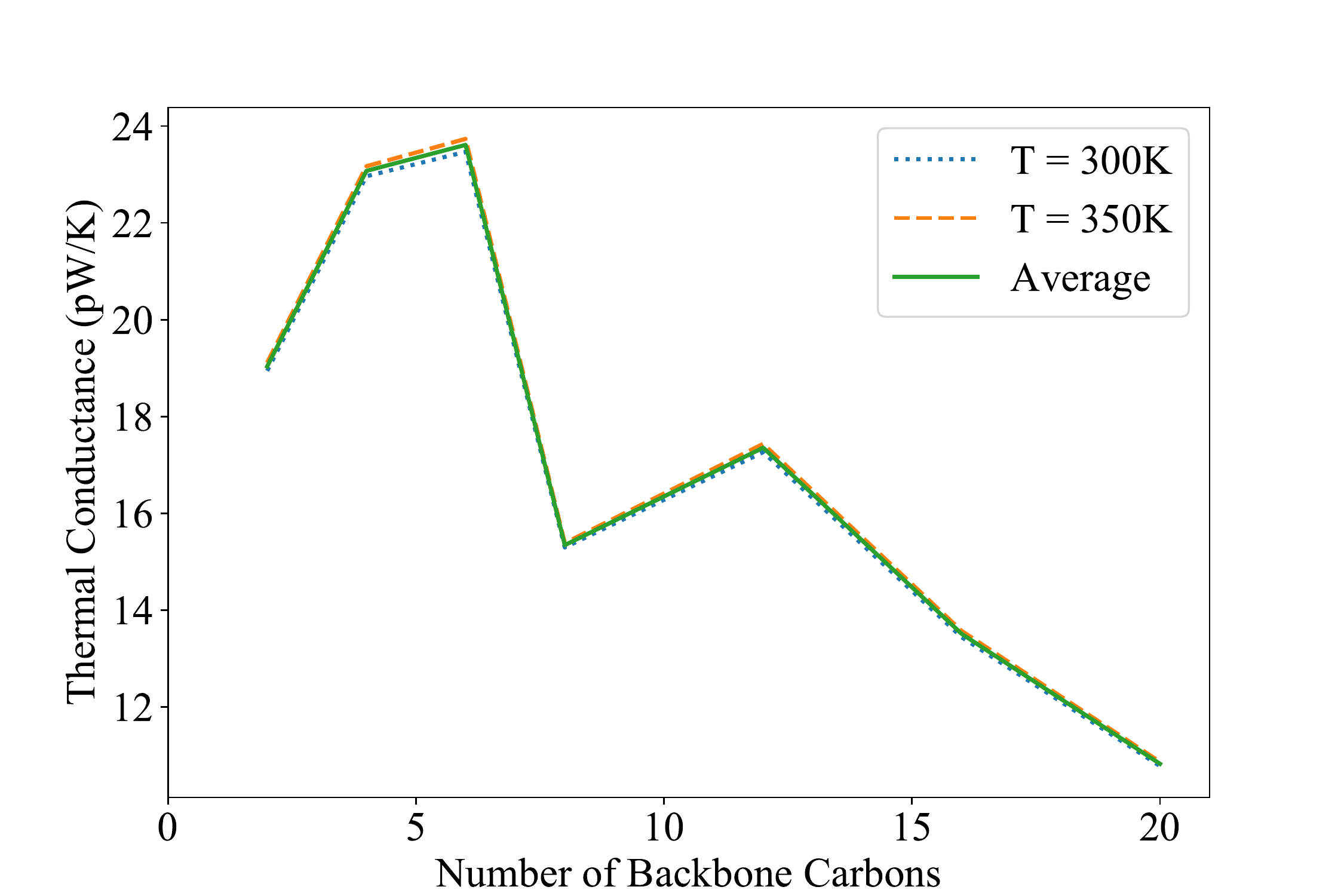}
    \caption{Comparison between derivative conductance and finite-bias conductance. For the finite-bias case, bath temperatures of 300K and 350K are taken (i.e. bias of 50K). This is compared with two derivative conductance calculations: Around a temperature of 300K and of 350K.}
    \label{fig:derivative_average_tempeture}
\end{figure}

We would now like to see how the choices of different conductance expressions (using the derivative form or the finite-bias form) influence the results.
It turns out that the influence is negligible across a range of alkane chain lengths (Figure \ref{fig:derivative_average_tempeture}). The conductance at 350K is slightly higher and 300K slightly lower, but the difference between either of them and the finite-bias result is significantly smaller than the values of the relevant conductances themselves.

\begin{figure}[H]
    \centering
    \includegraphics[width=1\textwidth]{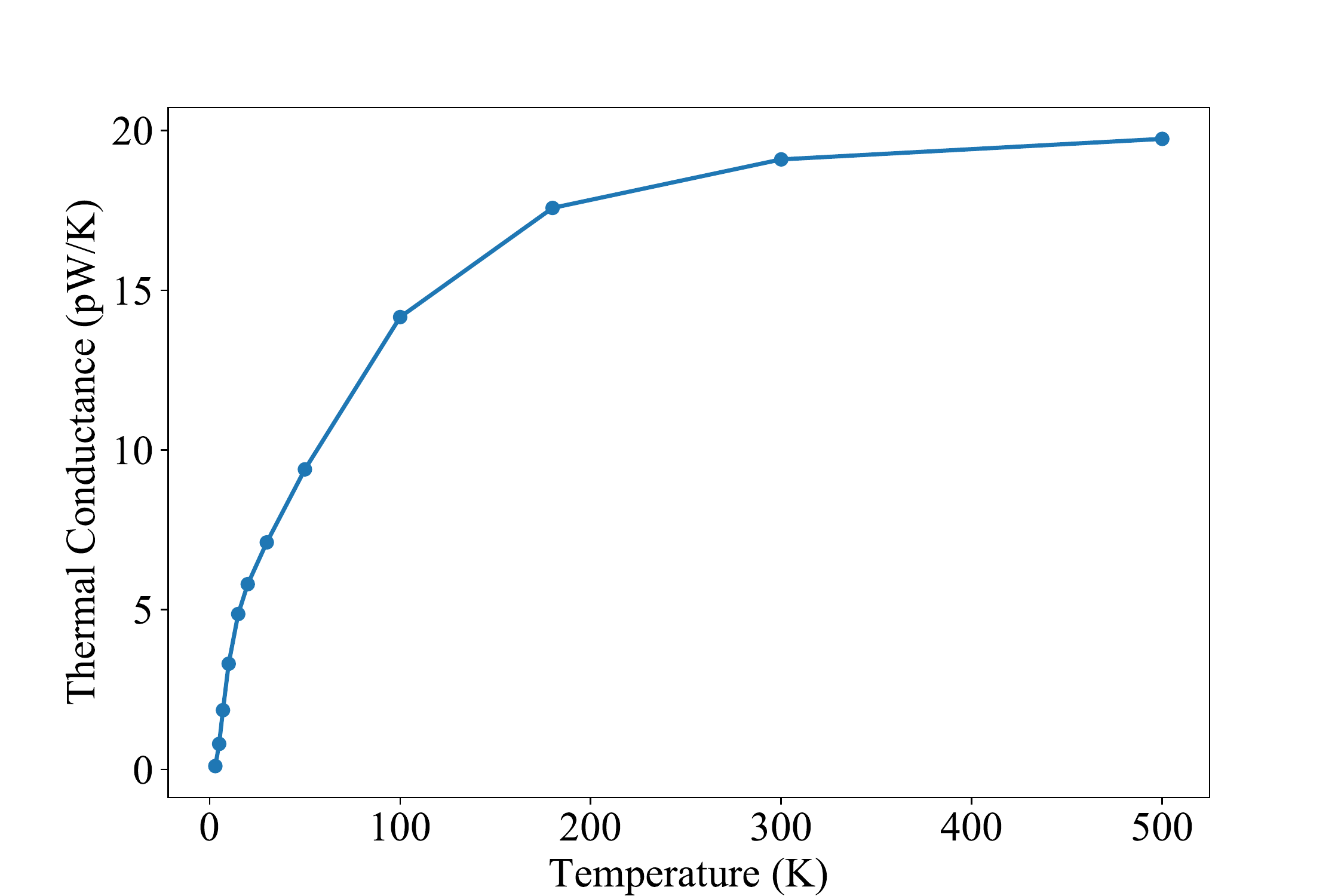}
    \caption{Temperature dependence of the derivative Landuaer thermal conductance, for butanedithiol (\ch{HS(CH_2)_10SH}) adsorbed on a substrate modelled by 3 layers of explicit Gold atoms.}
    \label{fig:temperature_Landauer_C10}
\end{figure}
Figure \ref{fig:temperature_Landauer_C10} shows the temperature dependence of the heat conductance of butanedithiol.
Conductance increases at a higher rate until 100K, and then starts to plateau as temperature increases further. This trend agrees with results reported for an identical alkane \cite{Kloeckner2016}.

\begin{figure}
    \centering
    \includegraphics[width=1\textwidth]{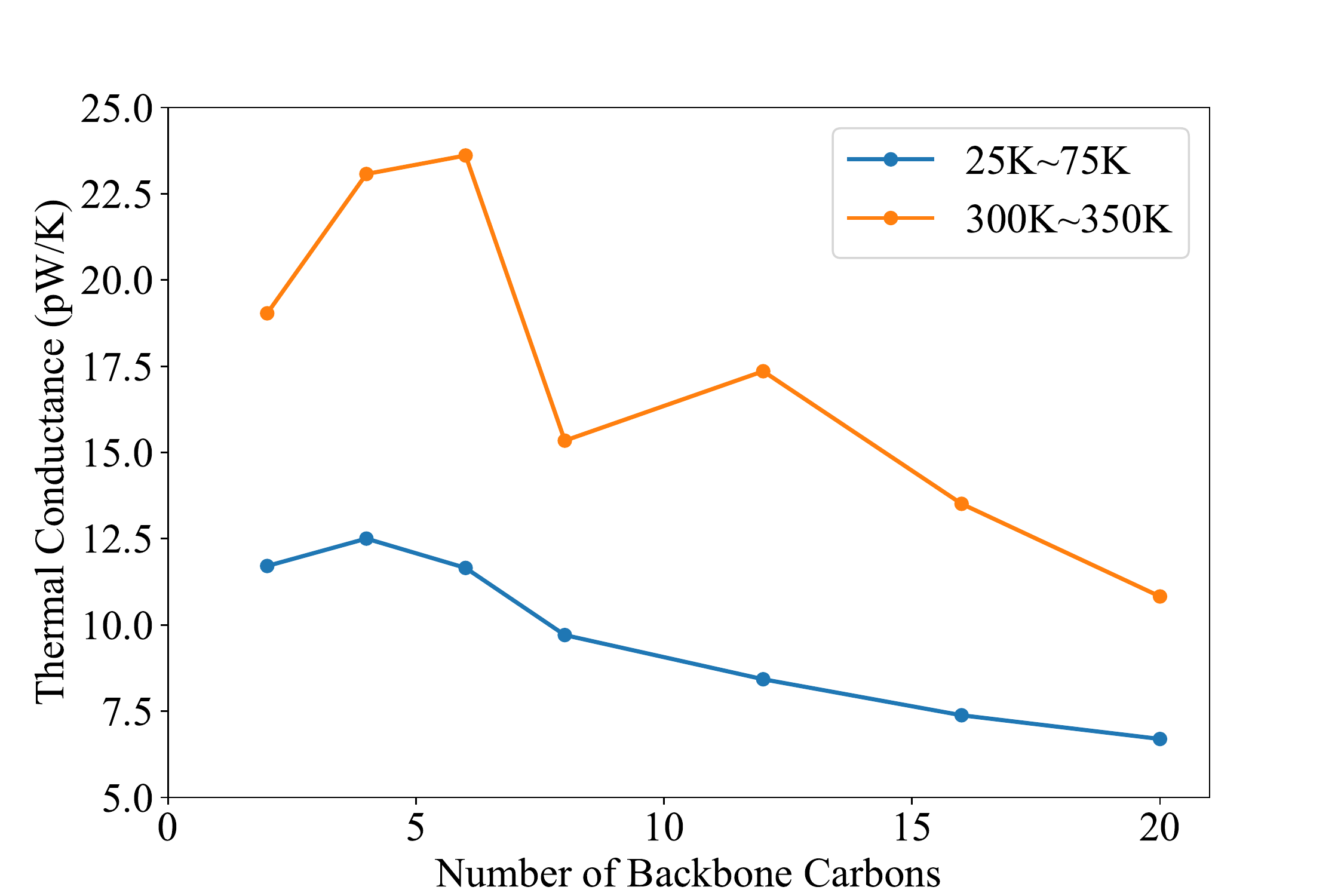}
    \caption{Comparison of Landauer conductance at low vs. room temperatures (temperature bias the same in both cases) for alkanedithiols of different lengths, with 3-layers of explicit gold bulks.}
    \label{fig:conductance_lnd_low_vs_room_temperature}
\end{figure}
The comparison of conductance between high (300K to 350K) and low (25K to 75K)(Figure \ref{fig:conductance_lnd_low_vs_room_temperature})
temperatures from Landauer calculations 
gives us a sense of the importance of the temperature factor 
plays in the molecular heat conduction simulations. 
From this perspective, the classical MD simulations (Figure \ref{fig:conductance_alkanes_lowTvsRoomT_upto16} is not a
reliable reference in the low temperature limit. 
The characteristic energy of phonons will be much lower, and thus the conductance at low temperature. The classical simulation does not show any major difference when comparing the high temperature profiles (Figure \ref{fig:conductance_alkanes_lowTvsRoomT_upto16}).
Therefore, we may propose a combination of probing heat conduction with classical MD at high temperature (where scattering might be more important), and with quantum Landauer's approach at low temperature limit (where the harmonic interaction dominates).
These benefits are all included in our newly developed simulation package presented here.

\bibliography{bib/j,bib/references,bib/achemso-demo}